\documentclass[reqno,12pt]{amsart}

\usepackage{amscd}
\usepackage{amssymb}
\usepackage{amsmath}
\usepackage{latexsym}
\usepackage[mathscr]{euscript}
\usepackage{epsfig}

\usepackage[usenames]{color}

\usepackage{fancybox}

\begin{document}

\numberwithin{equation}{section}

\newcommand{\bu}{\textcolor{blue}}
\newcommand{\red}{\textcolor{red}}
\newcommand{\mg}{\textcolor{magenta}}

\newtheorem{theorem}{Theorem}[section]
\newtheorem{definition}[theorem]{Definition}
\newtheorem{proposition}[theorem]{Proposition}
\newtheorem{lemma}[theorem]{Lemma}
\newtheorem{corollary}[theorem]{Corollary}
\newtheorem{rem}[theorem]{Remark}


\newcommand{\Aa}{{\mathcal A}}
\newcommand{\Bb}{{\mathcal B}}
\newcommand{\Cc}{{\mathcal C}}
\newcommand{\Dd}{{\mathcal D}}
\newcommand{\Ee}{{\mathcal E}}
\newcommand{\Ff}{{\mathcal F}}
\newcommand{\Gg}{{\mathcal G}}
\newcommand{\Hh}{{\mathcal H}}
\newcommand{\Ii}{{\mathcal I}}
\newcommand{\Jj}{{\mathcal J}}
\newcommand{\Kk}{{\mathcal K}}
\newcommand{\Ll}{{\mathcal L}}
\newcommand{\Mm}{{\mathcal M}}
\newcommand{\Nn}{{\mathcal N}}
\newcommand{\Oo}{{\mathcal O}}
\newcommand{\Pp}{{\mathcal P}}
\newcommand{\Qq}{{\mathcal Q}}
\newcommand{\Rr}{{\mathcal R}}
\newcommand{\Ss}{{\mathcal S}}
\newcommand{\Tt}{{\mathcal T}}
\newcommand{\Uu}{{\mathcal U}}
\newcommand{\Vv}{{\mathcal V}}
\newcommand{\Ww}{{\mathcal W}}
\newcommand{\Zz}{{\mathcal Z}}
\newcommand{\1}{{\mathcal I}}


\newcommand{\CC}{{\bf C}}
\newcommand{\HH}{{\bf H}}
\newcommand{\LL}{{\mathbf L}}
\newcommand{\NN}{{\bf N}}
\newcommand{\PP}{{\bf P}}
\newcommand{\RR}{{\bf R}}
\newcommand{\ZZ}{{\bf Z}}

\newcommand{\lL}{{\mathbf l}}
\newcommand{\mM}{{\mathbf m}}
\newcommand{\kaK}{{\mathbf \kappa}}
\newcommand{\id}{{\mathbf 1}}


\newcommand{\BM}{{\mathbb B}}
\newcommand{\CM}{{\mathbb C}}
\newcommand{\EM}{{\mathbb E}}
\newcommand{\GM}{{\mathbb G}}
\newcommand{\LM}{{\mathbb L}}
\newcommand{\NM}{{\mathbb N}}
\newcommand{\QM}{{\mathbb Q}}
\newcommand{\PM}{{\mathbb P}}
\newcommand{\RM}{{\mathbb R}}
\newcommand{\RMbar}{\overline{\mathbb R}}
\newcommand{\SM}{{\mathbb S}}
\newcommand{\TM}{{\mathbb T}}
\newcommand{\VM}{{\mathbb V}}
\newcommand{\ZM}{{\mathbb Z}}


\newcommand{\AG}{{\mathfrak A}}
\newcommand{\BG}{{\mathfrak B}}
\newcommand{\aG}{{\mathfrak a}}
\newcommand{\bG}{{\mathfrak b}}
\newcommand{\cG}{{\mathfrak c}}
\newcommand{\CG}{{\mathfrak C}}
\newcommand{\DG}{{\mathfrak D}}
\newcommand{\FG}{{\mathfrak F}}
\newcommand{\gG}{{\mathfrak g}}
\newcommand{\HG}{{\mathfrak H}}
\newcommand{\LG}{{\mathfrak L}}
\newcommand{\MG}{{\mathfrak M}}
\newcommand{\NNG}{{\mathfrak N}}
\newcommand{\PG}{{\mathfrak P}}
\newcommand{\SG}{{\mathfrak S}}
\newcommand{\TG}{{\mathfrak T}}
\newcommand{\WG}{{\mathfrak W}}
\newcommand{\Sssmall}{{\mathfrak s}}


\newcommand{\AV}{\vec{A}}
\newcommand{\aV}{\vec{a}}
\newcommand{\BV}{\vec{B}}
\newcommand{\bV}{\vec{b}}
\newcommand{\EV}{\vec{\Ee}}
\newcommand{\eV}{\vec{e}}
\newcommand{\fV}{\vec{f}}
\newcommand{\FV}{\vec{F}}
\newcommand{\gV}{\vec{g}}
\newcommand{\jV}{\vec{j}}
\newcommand{\JV}{\vec{J}}
\newcommand{\kV}{\vec{k}}
\newcommand{\LV}{\vec{L}}
\newcommand{\lV}{\vec{\ell}}
\newcommand{\MV}{\vec{M}}
\newcommand{\mV}{\vec{m}}
\newcommand{\nV}{\vec{n}}
\newcommand{\PV}{\vec{P}}
\newcommand{\pV}{\vec{p}}
\newcommand{\qV}{\vec{q}}
\newcommand{\rV}{\vec{r}}
\newcommand{\sV}{\vec{s}}
\newcommand{\SV}{\vec{S}}
\newcommand{\uV}{\vec{u}}
\newcommand{\vV}{\vec{v}}
\newcommand{\XV}{\vec{X}}
\newcommand{\xV}{\vec{x}}
\newcommand{\yV}{\vec{y}}
\newcommand{\naV}{\vec{\nabla}}
\newcommand{\muV}{\vec{\mu}}
\newcommand{\nuV}{\vec{\nu}}
\newcommand{\varphiV}{\vec{\varphi}}


\newcommand{\ha}{{\hat{a}}}
\newcommand{\hA}{{\hat{A}}}
\newcommand{\hq}{{\hat{q}}}
\newcommand{\hx}{{\widehat{x}}}
\newcommand{\hB}{{\widehat{B}}}
\newcommand{\hE}{{\widehat{E}}}
\newcommand{\hF}{{\widehat{F}}}
\newcommand{\hG}{{\widehat{G}}}
\newcommand{\hH}{{\widehat{H}}}
\newcommand{\hn}{{\widehat{n}}}
\newcommand{\hN}{{\widehat{N}}}
\newcommand{\hP}{{\widehat{P}}}
\newcommand{\hr}{{\widehat{r}}}
\newcommand{\hT}{{\widehat{T}}}
\newcommand{\hX}{{\widehat{X}}}
\newcommand{\hU}{{\widehat{U}}}
\newcommand{\hhX}{{\widehat{X}}}
\newcommand{\hW}{{\widehat{W}}}
\newcommand{\hDel}{{\widehat{\Delta}}}
\newcommand{\hgamma}{{\widehat{\gamma}}}
\newcommand{\hsig}{{\widehat{\sigma}}}
\newcommand{\hpi}{{\widehat{\pi}}}
\newcommand{\hDG}{{\widehat{\mathfrak D}}}


\newcommand{\as}{{\mathscr A}}
\newcommand{\bs}{{\mathscr B}}
\newcommand{\cs}{{\mathscr C}}
\newcommand{\ds}{{\mathscr D}}
\newcommand{\es}{{\mathscr E}}  
\newcommand{\fs}{{\mathscr F}}
\newcommand{\gs}{{\mathscr G}}
\newcommand{\hs}{{\mathscr H}}
\newcommand{\is}{{\mathscr I}}
\newcommand{\js}{{\mathscr J}}
\newcommand{\ks}{{\mathscr K}}
\newcommand{\ls}{{\mathscr L}}
\newcommand{\ms}{{\mathscr M}}
\newcommand{\ns}{{\mathscr N}}
\newcommand{\os}{{\mathscr O}}
\newcommand{\ps}{{\mathscr P}}
\newcommand{\qs}{{\mathscr Q}}
\newcommand{\rs}{{\mathscr R}}
\newcommand{\sss}{{\mathscr S}}
\newcommand{\ts}{{\mathscr T}}
\newcommand{\us}{{\mathscr U}}
\newcommand{\vs}{{\mathscr V}}
\newcommand{\ws}{{\mathscr W}}
\newcommand{\xs}{{\mathscr X}}
\newcommand{\ys}{{\mathscr Y}}
\newcommand{\zs}{{\mathscr Z}}


\newcommand{\ta}{{\tilde{a}}}
\newcommand{\tAa}{{\tilde{\Aa}}}
\newcommand{\tb}{{\tilde{b}}}
\newcommand{\tg}{{\tilde{g}}}
\newcommand{\tB}{{\tilde{B}}}
\newcommand{\tf}{{\tilde{f}}}
\newcommand{\tFf}{{\tilde{\Ff}}}
\newcommand{\txi}{{\tilde{\xi}}}


\newcommand{\Un}{{\underline{n}}}
\newcommand{\nU}{{\underline{n}}}
\newcommand{\uU}{{\underline{u}}}


\newcommand{\ccd}{{\c c}}
\newcommand{\ii}{{\^\i}}
\newcommand{\itr}{{\"\i}}
\newcommand{\eme}{$^{\grave eme}$ }


\newcommand{\TR}{{\rm Tr\,}}                       
\newcommand{\tr}{{\rm tr\,}}                       
\newcommand{\TV}{\Tt_{\PM}}                        
\newcommand{\tra}{\mbox{\sc t}}                    
\newcommand{\tri}{\mbox{\sc\tiny t}}               
\newcommand{\Cs}{$C^{\ast}$-algebra }              
\newcommand{\CS}{$C^{\ast}$-algebra}               
\newcommand{\Css}{$C^{\ast}$-algebras }            
\newcommand{\CsS}{$C^{\ast}$-algebras}             
\newcommand{\AGl}{\AG_{loc}}                       
\def\essinf{\mathop{\rm essinf}}                   
\def\esssup{\mathop{\rm esssup}}                   
\newcommand{\bc}{b^{\dag} }                        
\newcommand{\ba}{b}                                
\newcommand{\Kki}{{\mathcal K}_{\infty}}           
\newcommand{\req}{\rho_{\mbox{\scriptsize\em eq}}} 
\newcommand{\eq}{_{\mbox{\scriptsize\em eq}}}      
\newcommand{\gns}{\pi_{\mbox{\scriptsize\sc gns}}} 
\newcommand{\Sp}{\mbox{\rm Sp}}                    
\newcommand{\Ker}{\mbox{\rm Ker}}                  
\newcommand{\Prob}{\mbox{\rm Prob}}                
\newcommand{\supp}{\mbox{\rm supp}}                
\newcommand{\supi}{\mbox{\rm \footnotesize supp}}  
\newcommand{\gau}{\gamma}                          
\newcommand{\gao}{\overline{\gamma}}               
\newcommand{\CP}{\mbox{\rm CP}}                    
\newcommand{\HS}{\mbox{\rm HS}}                    
\newcommand{\diam}{\mbox{\rm diam}}                
\newcommand{\diami}{\mbox{\rm \tiny diam}}         
\newcommand{\eff}{\mbox{\it \tiny eff}}            


\newcommand{\kB }{k_{\mbox{\tiny \it B}}}          
\newcommand{\omet}{\tilde{\omega}}                 
\newcommand{\rav}{r_{av}}                          
\newcommand{\nel }{n_{\mbox{\tiny \it el}}}        
\newcommand{\rB }{a_{\mbox{\tiny \it B}}}          
\newcommand{\jexp}{j_{\mbox{\tiny \it exp}}}       
\newcommand{\sig}{\underline{\sigma}}              


\title{Dissipative dynamics in semiconductors at low temperature}


\author{George Androulakis}
\address[Androulakis]{Department of Mathematics, University of South Carolina, Columbia, SC 29208}
\email{giorgis@math.sc.edu}

\author{Jean Bellissard}
\address[Bellissard]{Georgia Institute of Technology, School of Mathematics, Atlanta GA 30332-0160}
\email{jeanbel@math.gatech.edu}

\author{Christian Sadel}
\address[Sadel]{Department of Mathematics, University of California Irvine, Irvine, CA 92697-3875}
\email{csadel@math.uci.edu}

\thanks{J. Bellissard and C. Sadel were supported by NSF grants DMS-0600956 and DMS-0901514.}

\subjclass{82B10, 81Q10, 70F45, 47D07, 46N50}

\date{7/5/2011}

\begin{abstract}
A mathematical model is introduced which  describes the dissipation of electrons in lightly doped semi-conductors. 
The dissipation operator is proved to be densely defined and positive and to generate a Markov semigroup of operators.
The spectrum of the dissipation operator is studied and it is shown that zero is a simple eigenvalue, which makes the 
equilibrium state unique. Also it is shown that there is a gap between zero and the rest of its spectrum which makes 
the return to equilibrium exponentially fast in time.
\end{abstract}

\maketitle

\vspace{.5cm}

\noindent {\footnotesize{\em The sciences do not try to explain, they hardly even try to interpret, they mainly make models. 
By a model is meant a mathematical construct which, with the addition of certain verbal interpretations, describes observed phenomena. 
The justification of such a mathematical construct is solely and precisely that it is expected to work.} (J. von Neumann \cite{vN})}


\section{Introduction}
\label{mott09.sec-intro}

\vspace{.3cm}

\noindent This article is dedicated to the construction and the fundamental properties of a model of dissipative transport, describing the 
electron or hole transport in semiconductors at very low temperature. By ``very low'' it is meant that the temperature is low enough so as 
to confine the charge carriers to the impurity band. Without dissipation, the transport is {\em coherent} and is likely to be described by an 
Anderson model, namely a Schr\"odinger operator on a discrete lattice with a random potential. In {\em lightly doped} semiconductors this 
model has to be considered in the {\em strong localization regime} as will be explained in Section~\ref{mott09.sec-phys}. The main source of 
dissipation in all solids, including semiconductors, is coming from the {\em electron-phonon} interaction. Namely, the Coulomb interaction 
between electrons and nuclei leads to the slow nuclei motion when an electron is passing by. The harmonic interaction between nuclei leads, 
in turn, to the production of acoustic waves, that are quantized, at least if produced in small quantities. These quanta of acoustic waves 
are called {\em phonons}. These waves propagate in the crystal at the speed of sound and can kick another electron eventually, leading to 
loss of information about the electron quantum state. There are other sources of dissipation for the electron motion like the direct Coulomb 
interaction between them, spin coupling or photon emission. However, it has been shown \cite{Mahan} that the first process is quantitatively 
more important than all others. Still, even the dominant mechanism for dissipation can be considered as weak in most cases, in particular 
for the problem considered in the present paper. Therefore, it is legitimate to approximate the system by using the so-called 
{\em Markov approximation}. As a consequence, the dissipative dynamic can be described through a Markov semigroup whose generator  
is a Lindblad operator \cite{Li76}, also called nowadays {\em Lindbladian}. Like in Atomic Physics \cite{CTDRG04}, this Lindbladian could be 
computed from their second order perturbation theory, called the {\em Fermi golden rule}. It will not be the method used here. The purpose of 
the present work will be to construct a Lindblad operator that describes phenomenologically the dominant sources of dissipation, and to 
investigate its spectral properties. The precise description
of the model is given in Subsection~\ref{mott09.ssect-model}. The spectral properties are summarized in Section~\ref{mott09.sect-model}. The 
main result is the Theorem~\ref{mott09.th-spectDiss}, showing that (i) such an operator and the dynamical semigroup it generates are well 
defined even if the system is out of equilibrium and that (ii) it forces the return to equilibrium if there is no gradient of chemical 
potential or of temperature in the system. However, the main new contribution of the present work lies in the mathematical framework as 
explained below. But since the explicit model is strongly dependent of the physical regime that it intends to describe, it will be necessary 
to describe the physics in detail in order to make sure that the model is realistic (see Section~\ref{mott09.sec-phys}).

\vspace{.2cm}

\noindent The problem investigated here is motivated by the mechanism called {\em variable range hopping conductivity} \cite{ES}. It has 
dominated the study of semiconductors for almost two decades since the work of Miller and Abrahams in 1960 \cite{MA60} and the seminal 
contribution of Mott \cite{Mott} predicting the behavior of the conductivity as a function of the temperature. It has been 
suggested \cite{PS93} that this regime is also the dominant contribution in the Quantum Hall Effect \cite{BSBvE} that explains in particular 
the amazing accuracy of the experiment. For indeed, the relative error with which the Hall conductivity can be measured in a QHE experiment 
is of the order of $10^{-10}$ due to the smallness of the direct conductivity on the plateau of conductivity \cite{PG}. The theory of the 
integer QHE has been made rigorous through the use of the formalism of Non-commutative Geometry in \cite{BSBvE}. In this latter work, the 
problem of dissipative transport was investigated within the so-called {\em relaxation time approximation} (RTA). The RTA reduces the 
dissipative mechanism to only one time scale and allows to consider the charge carriers as independent particles. As shown in \cite{BSBvE}, 
when applied to the QHE, this approximation leads to a relative error of $10^{-4}$ with the best data, namely six orders of magnitude larger 
than what is actually observed~!! The reason is that the charge carriers conductivity is suppressed by the variable range hopping, as was 
shown by Mott \cite{Mott}. However, such a mechanism involves an infinite number of time scales. It was advocated in \cite{Be03} that such 
a mechanism can be represented through a Markov semigroup, the generator of which is a Lindblad operator. Unfortunately previous attempts 
to implement this idea have provided mixed results. The main reason is that the charge carriers, electrons or holes, are Fermions and any 
approximation leading to consider these particles in a semi classical regime fails to include the statistical correlation induced by the 
electron-phonon interactions. In order to successfully represent this mechanism, a mathematical model must take second quantization into 
account. The main new contribution of the present paper is precisely to work with a many-body formalism. Since this approach is technically 
very demanding, the model will be simplified. The main simplification consists in replacing the Anderson model, describing the coherent part 
of the motion, by a purely potential contribution, neglecting the kinetic part, which, in real semiconductors is indeed extremely small. 
This kinetic part will be reintroduced in the dissipative mechanism through a contribution of the tunneling effect between impurities.

\vspace{.2cm}

\noindent Even with so many simplifications, the formalism is heavy and will occupy most of this paper. This is because the random character 
of the distribution of impurities breaks the translation invariance. Since the early eighties, thanks to using the ideas of Non-commutative 
Geometry \cite{Co94}, the formalism required to replace the translation group is known (for instance, see \cite{Be03} and references therein): 
a groupoid replaces the group of translations. The inclusion of the many-body formalism in this framework was developed in the PhD Thesis of 
Dominique Spehner \cite{Sp00}. This leads to replace the observable algebra by a {\em continuous field} of \Css over the space describing the 
disorder. The notion of continuous field of Banach spaces was introduced by Tomyama \cite{Ty58,Ty62,Ty63} in the context of the spectral 
theory for \Css and later developed by Dixmier \cite{Dix}. While the concept is easy to understand, it is technically demanding. Then, the 
coherent dynamics can be defined as a continuous field of dynamics, leading to a continuous field of KMS-states describing the equilibrium 
of the electron gas in the solid. In much the same way, the dissipative dynamics is defined by a continuous field of Markov semigroups. In 
the present work, various  existence results for the dynamics are proposed. One, mainly the Theorem~\ref{mott09.th-convAG}, is based on the 
estimates used by Bratteli and Robinson \cite{BrRo2} to prove the existence of the dynamics in the many-body theory. The other one, mainly 
the Theorem~\ref{mott09.th-close}, uses the continuous field of Hilbert spaces generated by the field of KMS-states, through the 
GNS-construction, and sees the Lindbladian as a non-commutative analog of a {\em Dirichlet form}. Dirichlet forms were introduced by 
Beurling and Deny \cite{BeDe59} and the characterization was completed by
Fukushima \cite{Fuk}. 
The noncommutative Dirichlet forms were defined by Albeverio and Hoegh-Kr\o hn \cite{AlHK77} and they were characterized in full generality by Cipriani \cite{Ci97}. The definition requires some notion of positivity in the Hilbert space of the GNS-representation. 
Such a positivity is provided by a cone in the Hilbert space, that was identified and characterized in the early seventies by Araki \cite{Ar74} and Connes \cite{Co74}. In each fiber of the field of Hilbert spaces, it is the cone generated by positive elements of the corresponding fiber of 
the field of observable algebras. For the sake of the reader, this will be explained in Section~\ref{mott09.ssect-gns}.

\vspace{.2cm}

\noindent Because the formalism required here is so heavy, it seems wiser to restrict the present paper to the description of the dynamics 
and the return to equilibrium. However, the real goal is to show that this model is liable to account for the Mott prediction concerning the 
low temperature behavior of the conductivity. An important result was obtained by \cite{FSBS}: by looking at the variable range hopping at 
very large length scale, as a random walk in a random environment the authors could prove that the Mott prediction was a lower bound to the 
diffusion constant. However, to get an upper bound is highly non trivial.
In order to do so, using the present model, it will be necessary to face two challenges. The first one is the definition of the local 
currents. As it turns out, this is not a trivial problem because of possible divergent effects in the infinite volume limit. It requires 
insight about the physical nature of currents and of the dissipation mechanisms. With the proper concept, though it is possible to prove 
rigorously the validity of the Kubo formula \cite{ABS11} whenever the charge carriers can be considered as a continuous fluid. The other 
challenge is the discontinuous nature of the variable range hopping mechanism, forcing the charge carriers to hop at distances ten times 
larger than the average distance between impurities. In particular, the {\em continuous fluid picture breaks down}~! This is why Miller 
and Abrahams \cite{MA60} proposed to see the charge carriers as electric currents in a random network. In the early seventies percolation 
theory was successfully introduced into this picture \cite{AHL71,Po72,ES} to justify the prediction of Mott. However, several approximations, 
justified by the physical situation, require additional work in order to make this argument rigorous within the scope of the present model. 
It will be hopefully the subject of a forthcoming publication.

\vspace{.3cm}

\noindent {\bf Acknowledgments: } This work benefited from the NSF grants  DMS-0600956 and DMS-0901514. Part of this work was done in 
Bielefeld with the support of the SFB 701 ``Spectral Structures and Topological Methods in Mathematics'' during the Summers 2009 and 2010. 
G.A. and C.S. thanks the School of Mathematics at the Georgia Institute of Technology for support during the Spring 2009.


\section{Physics of lightly doped semi-conductors}
\label{mott09.sec-phys}  

 \subsection{Orders of Magnitude}
 \label{mott09.ssect-ord}

\noindent The content of this section can be found in several textbook, in particular the one by Shklovskii and Efros \cite{ES}.

\vspace{.1cm}

\noindent The two types of semi-conductors that are the most used and studied today are {\em silicon} and $GaAs$, 
due to their importance in modern electronics, telecommunication and computer hardware. Silicon is currently obtained, in the industrial process, in a 
form of cylindrical ingots of about $2\,m$ in length and $25\,cm$ in diameter. The crystal is perfect with less than $10^{-10}$ impurity 
or defect per atom. Because silicon has $4$ valence electrons, the atomic orbitals hybridate in the $sp_3$ form, leading  to a diamond crystal, 
where tetrahedra alternating in a staggered way. $Ga$ belongs to the column III of the periodic table, namely it has $3$ valence electrons. 
$As$ has $5$ valence electrons and thus, it belongs to the column V. Mixed together in equal quantity, $Ga$ and $As$ exchange one electron 
to produce pairs of tetrahedra, leading also to a diamond lattice. If the electron-electron interactions are ignored, the band spectrum is 
similar to the one of graphene, namely two bands touch on a conical point exactly at the Fermi level, (the Fermi level is the maximum energy of an electron at zero temperature).

\begin{figure}[htbp]
\centerline{\includegraphics*[clip=true,width=110mm]{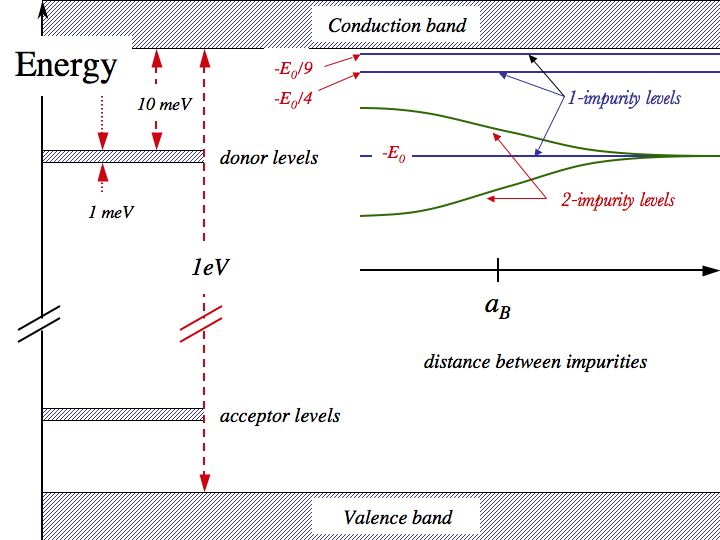}}
  \caption{(i) Left: energy ranges (ii) Right: 1 \& 2 hydrogen atom levels}
  \label{mott09.fig-band}
\end{figure}
\vspace{.1cm}

\noindent However, the Coulomb interaction between the valence electrons belonging to the same atom, is strong enough to force the 
opening of a gap at the Fermi level, which will be called the {\em main gap}. For $Si$, this gap is about $1.1\,eV$, which is enormous. 
Correspondingly, the temperature necessary to allow a large number of electrons to jump from the valence band to the conduction band 
would be $1.2\times 10^4\,K$. In  other words, the clean silicon is a perfect insulator. For $GaAs$ the gap is $1.43\,eV$ corresponding 
to a temperature of $1.6\times 10^4\,K$.

\vspace{.1cm}

\noindent For the purpose of electronic applications, the doping is about $10^{-9}$ impurity per atomic site. Such a crystal is called 
{\em lightly doped}. For $n$-type doping, the impurities are atoms with one more electron in the valence band than the crystal atoms (such as $P,As,Sb$). 
Then, an impurity band is created in the gap at a distance of about $10\,meV$ from the conduction band (see Fig.~\ref{mott09.fig-band}). 
Hence, the gap between the impurity band and the conduction band is $100$ times smaller than the main gap. In particular, at room temperature, 
most electrons of the impurity band jump into the conduction band, which explains why semi-conductors are actually conductors. However, 
the conduction electron density is controlled by the impurity concentration, namely it is much smaller than in metals. At very low temperatures, 
say around  $1\,K$, this gap is too large to allow electrons to jump, so that electrons are confined in the impurity band (see Fig.~\ref{mott09.fig-band}). 
Similarly, for a $p$-type doping, the impurities are acceptors instead, namely they have one electron less on their valence band than the 
crystal atoms (such as $Al,Ga,In$). Then, a hole-impurity band is created near the valence band at approximately the same distance as before. 
Similar conclusions arise after proceeding to the {\em hole-particle symmetry}: (i) the charge carriers are holes instead of electrons, 
(ii) the origin of the energy is the top of the valence band instead of the bottom of the conduction band and (iii) the sign of energy is reversed.

\vspace{.1cm}

\noindent For an impurity concentration $c\sim 10^{-9}$ per atom, the average distance between impurities is given by 
$\rav = c^{-1/3}\sim 1000$ atomic spacings. As will be seen in the next Section~\ref{mott09.ssect-hyd}, an isolated impurity behaves 
like an Hydrogen atom with a re-normalized mass and a re-normalized Coulomb coupling constant. In these materials, the Bohr radius of an impurity is 
$\rB \simeq 100$\AA. Hence the average distance between impurities corresponds approximately to $10 \rB$. This gives a band width of 
approximately $W\simeq 1\,meV$, $10$ times smaller than the gap between the impurity band and the conduction (or valence) band (see Fig.~\ref{mott09.fig-band}).

\vspace{.1cm}

\noindent In most cases a semi-conductor has the two types of impurities simultaneously. In such a case it is called {\em compensate}. 
If $n$-type impurities dominate, then all acceptors will get an extra electron since their energy is lower, creating negative ions, 
and some proportion of the donors will become ionized creating positive charges in the crystal. The Fermi level, namely the maximum 
energy of an electron at zero temperature, will be located within the donor impurity band at a position depending upon the relative 
concentration of donor and acceptor. If the acceptors dominate, some of them will acquire all the electrons coming form the donors 
and will become negatively charged, while all the donors will be ionized. The Fermi level will then be inside the acceptor impurity band 
and the conduction will be due to holes instead. In both cases however, the 
positions of the ions are random, 
creating a random Coulomb field within the crystal which influence the energy level of the impurity electrons.

 \subsection{One Impurity \& Hydrogen Atom}
 \label{mott09.ssect-hyd}

\noindent If one donor impurity is inserted in a perfectly periodic crystal, Slater \cite{Sl49} showed in 1949
using a theorem by Wannier \cite{Wa37}, 
that the extra electron behaves like in an hydrogen atom, with re-normalized parameters for its mass and the dielectric constant, 
provided the origin of energies is taken at the bottom of the conduction band. If the impurity is an acceptor instead, a similar argument 
can be used for holes instead of electrons, using electron-hole symmetry. If the bottom of the conduction band is an isotropic minimum, the corresponding effective Hamiltonian describing the electron near the impurity is given by
\begin{equation}
\label{mott09.eq-1H}
H_H = -2E_0
      \left(
       \frac{\Delta_r}{2} + \frac{1}{|r|}
      \right)\,,
\hspace{1cm} 
r=\frac{x}{\rB}\,,\;\;
 \rB = \frac{\hbar^2 \kappa}{me^2}\,,\;\;
  E_0 = \frac{me^4}{2\hbar^2\kappa^2}\,.
\end{equation}
\noindent Here, $e$ is the electron charge, $m$ is the effective mass of the electron in the crystal and $\kappa$ is the re-normalized Coulomb 
constant in the medium. The ground state is given by 

\begin{equation}
\label{mott09.eq-gsH}
\phi_0(x) =
   \frac{e^{-|x|/a_B}}{Z_0}\;,
\qquad
Z_0=\left(\pi \rB^3 \right)^{1/2}
\end{equation}

\noindent so that $\rB$, called the {\em Bohr radius}, gives the length scale beyond which the wave function becomes negligible. 
The corresponding eigenvalue is $-E_0$ given in eq.~(\ref{mott09.eq-1H}). The other energy levels are given by $E_n =-E_0/n^2$ for
 $n=1,2,3,\cdots$. In particular the gap between the ground-state and the first excited state is $3E_0/4$ (see Fig.~\ref{mott09.fig-band}).

\vspace{.1cm}

\noindent As a result, in the donor case, the lower energy state of the impurity is located in the gap of the crystal, at a distance $E_0$ from the conduction band. Applied to impurities in a crystal, the Wannier equivalence theorem provides effective values for $m$ and $\kappa$, which are usually different from the values for the electron in the vacuum. For silicon, this Bohr radius is about $100${\AA} and $E_0 \simeq 10\,meV$ \cite{ES}.

 \subsection{The Anderson model}
 \label{mott09.ssect-densimp}

\noindent In a semi-conductor, there is a density of impurities. Even if this density is as small as $10^{-9}$, as it is for lightly doped media, the effect on the charge carrier dynamics is not negligible. The first important effect occurs if the semi-conductor is compensated. For indeed since the density of donors and acceptors are not equal, there is a nonzero density of ions in the crystal. The position of these ions is random. Consequently, they create a static electric field within the crystal, which is also random. Since the charge carriers are confined on the impurity site, they cannot move to screen this field. As a consequence, the effective potential seen by each charge carrier at an impurity site, is itself shifted by a random term. Since the distance between ions is very large, the values of potential at each impurity site might be considered as 
independent random variables. By homogeneity, they should have the same distribution.

\vspace{.1cm}

\noindent On the other hand, since the ground state wave function decays exponentially fast, the tunneling of a charge carrier between two impurity sites is controlled by the overlap of the two wave packets localized on each of the impurities. This overlap is of the order of $e^{-r/a_B}$ if $r$ is the distance between the two impurities. If the average distance between impurities is about $10\rB$, this term is very small (of the order of $10^{-4}$). This implies that the effective Hamiltonian describing the motion of a charge carrier through the impurity sites is made of two contributions: (i) the contribution of the random potential created by the ions in the crystal and (ii) the hopping term of the order of $e^{-r/a_B}$. Since this latter contribution is much smaller than the potential contribution, this effective Hamiltonian is given by an Anderson model in which the kinetic term is much smaller than the potential term. Hence, {\em paradoxically, lightly doped semi-conductors, at very low temperature, correspond to a strong disordered Anderson model}. In the present paper, the hopping term will be ignored, so that the one-particle Hamiltonian will be a pure on-site potential.



\subsection{Mott's Variable Range Hopping Transport}
 \label{mott09.ssect-mott}

\noindent In 1968, Mott \cite{Mott} proposed the following argument liable to explain the conductivity properties of Anderson insulators at low temperature (see also \cite{ES}). Mott assumes that the solid is a $d$-dimensional Anderson insulator, namely the electrons are strongly localized. In particular, each energy level of the electron energy spectrum is associated with a position in the solid within a ball of diameter given by the localization length $\xi$ (which can be taken of the order of $\rB$). Moreover, he assumes that the density of states (DOS) $n_F$ at the Fermi level is non  vanishing but small enough so as to avoid overlap between electron states. In particular the mean distance between neighboring electron states is large compared to $\xi$. The inverse of $n_F\cdot \xi^d$ is a measure of the mean level spacing between states within a ball of diameter $\xi$. The temperature will be small compared with this scale, namely
$$
\kB  T n_F \xi^d \;=\; \frac{T}{T_0}\;\ll \; 1\,,
$$
\noindent if $\kB $ denotes the Boltzmann constant. Then within a small error, all states with energy 
smaller than the Fermi level $E_F$ are occupied, whereas the ones with higher energy are empty. The probability that a phonon of energy $\varepsilon$ is produced is proportional to $e^{\varepsilon/\kB T}$. This is true provided $\varepsilon \gg \kB  T$. 
Such a phonon can be absorbed by an electron of the Fermi sea localized at $x\in\Ll$, with energy $\epsilon_x < E_F$, and cause
this electron to hop into a state localized at $y\in\Ll$ with energy $\epsilon_y > E_F$.
The probability for such an event to happen is controlled by the tunneling effect forcing the electron to move from $x$ to $y$. Let $r=|x-y|$ be the distance between such states, so that the tunneling probability be proportional to $e^{-r/\xi}$. Therefore, the probability $P$ of transfer of an electron at distance $r$ from its original location is proportional to
\begin{equation}
\label{mott09.eq-probmot}
P \; \propto \; 
   \exp\left[-\left\{
           \frac{\varepsilon}{\kB  T} +
            \frac{r}{\xi}
         \right\}
        \right]
\end{equation}
\noindent  By definition of the DOS, the product $\varepsilon\cdot n_F\cdot r^d$ represents the average number of states in an energy interval of width $\varepsilon$ localized in a cube of size $r$. Mott then argued that the most likely value of the radius satisfies 
\begin{equation}
\label{mott09.eq-nFbound}
n_F\cdot\varepsilon \cdot r^d \; \approx \; 1\,.
\end{equation}
\noindent The probability of jump $P$ is then optimized over $r$ and $\varepsilon$. This can be done by giving $r$ its minimum value compatible with eq.~(\ref{mott09.eq-nFbound}) and by maximizing {\em w.r.t.} $\varepsilon$. Therefore the conductivity, which is proportional to the sum of contributions of all such jumps, will be controlled by the maximal value of $P$. Optimizing over $\varepsilon$ leads to
\begin{equation}
\label{mott09.eq-mottlaw}
P \; \propto \; 
    \exp\left[-\left\{
           \frac{T_0}{T}
         \right\}^{1/(d+1)}
        \right]\,,
\qquad
\kB T_0 = \frac{(d+1)^{d+1}}{d^d}\frac{1}{n_F \xi^d}\,. 
\end{equation}
\noindent Then, the phonon energy optimizing $P$ is given by $\varepsilon_{opt}\approx T^{d/d+1} \gg T$, while the average distance of the jump is given by $r/\xi \approx (T_0/T)^{1/d+1} \gg 1$.

\vspace{.1cm} 

\noindent Clearly, the conductivity is proportional to the probability of transfer of electrons per unit time. Therefore, we expect the conductivity to be proportional to the same factor as a function of the temperature. 

\vspace{.1cm} 

\noindent For lightly doped $3D$-semiconductors like silicon (see 
\cite{ES}), the localization length is given by the Bohr radius of the 
impurity, which is about $100$\AA. For a concentration of $10^{-9}$, the mean distance between impurities is $1000${\AA} which is approximately $10 \xi$. Then, the typical width of the impurity band is about $1 meV$. Assuming the DOS to be flat on it gives $n_f=c/W$ which leads to 
$T_0 \approx 1.1\times 10^5\,K$~! This is huge indeed. At $T=1\,K$, this gives $(T_0/T)^{1/4}\simeq 18$, namely (i) the electron hops at about $18\xi\simeq 2\rav$, (ii) the conductivity is multiplied by a factor $e^{-18}\simeq 1.2\times 10^{-8}$ due to this mechanism~! It turns out that this is exactly what happens in the quantum Hall effect (QHE) \cite{PG,BSBvE}: the Mott variable range hopping controls the fluctuation of the plateaus, leading to the amazing accuracy
of this experiment.

\vspace{.1cm}

\noindent Mott's law has been well documented in the experimental physics literature. A review of the results obtained in the seventies on various semi-conductors can be found in \cite{Hi76} (see also \cite{Za77}) and some results have been reported in \cite{MPPW}. In addition a large part of the book by Efros and Shklovskii \cite{ES} is dedicated to this effect (see Chapter 9, in particular and references therein) .

\vspace{.1cm}

\noindent The previous version of Mott's argument is based on a critical assumption summarized by eq.~(\ref{mott09.eq-nFbound}). This part of the argument has been the focus of attention of several works in the early seventies \cite{Zi68,EC70,AHL71,Ki71,LT71,Po72,Be73} and it is the main topic of a large fraction of the book of Shklovskii and Efros \cite{ES}. Following the description provided by Miller and Abrahams \cite{MA60}, the electron conduction is seen as a random resistor network. In the early sixties, Ziman \cite{Zi68} suggested that percolation theory should be a key technique to investigate hopping transport. Since then it became indeed increasingly clear that percolation gives the right argument to justify eq.~(\ref{mott09.eq-nFbound}) and Mott's prediction.


 \subsection{Description of the Model}
 \label{mott09.ssect-model}

\noindent Based upon the argument of Mott, it becomes possible to propose a model. This section will be dedicated to its heuristic description, leaving the rigorous definitions for the Sections that follow. All along the present paper, only $n$-type doped semiconductors will be considered. A particle-hole symmetry permits to consider the $p$-type case.

\vspace{.2cm}

\noindent Let $\Ll$ be the lattice $\ZM^d$. For any $x \in \Ll$ let $s_x \in \{ 0, 1\}$ be a random variable
with $s_x=1$ if and only if an electron state is available at $x$. Given a family of random variables 
$s=(s_x)_{x \in \Ll}$ each taking values in $\{ 0, 1 \}$, let $\Ll (s)$ be the random subset of $\Ll$ containing
all sites $x$ where $s_x=1$.
Thus $\Ll (s)$ is the set of sites in the semiconductor on which an impurity electron state is available. 
In the tradition of Solid State Physics, and only for the heuristic description of the model, the total number of sites available will be considered as finite. The electron will be described in the second quantization picture, (see \cite{Da76}), through a pair of Fermion creation-annihilation operators $(a_x^\dag, a_x)$, where $x\in\Ll(s)$ is a lattice site, thus obeying the 
canonical anti-commutation rules which we abreviat by CAR,
\begin{equation}
\label{mott09.eq-car}
a_x a_y+a_ya_x = 0\,,
\quad
   a_x^\ast = a_x^\dag\,,
\quad
    a_xa_y^\dag+a_y^\dag a_x = \delta_{xy}\id \,.
\qquad \mbox{\rm \bf (CAR)}
\end{equation}
\noindent Here $A^\ast$ denotes the adjoint of the operator $A$. In this work, for simplicity, only spineless electrons are considered, because the model proposed will not couple the spin to the motion. The quantum dynamics will be made of two parts: (i) the coherent part, describing the electron motion in absence of dissipation, and (ii) the dissipative part, taking into account the interaction with other degrees of freedom, especially the phonons. 

\vspace{.1cm}

\noindent Since the dynamic concerns the electron gas, the system will be considered in the {\em local equilibrium approximation} \cite{Be03}. Namely the electron gas is seen as the union of {\em mesoscopic cells} with size large enough so that within a {\em mesoscopic time} each such cell has the time to return to equilibrium. On the other hand the size of the cells is small compared to the size of the sample under consideration. In each mesoscopic cell, the electron gas is in an equilibrium state, namely defined by the usual thermodynamic parameters, the temperature $T$ (or rather the inverse temperature $\beta= 1/\kB T$) and the chemical potential $\mu$. Such mesoscopic cells are opened, namely they allow both energy and electrons to be exchanged from cell to cell. Therefore, only the average of the energy and of the particle number is fixed. In addition, if the system is put out of equilibrium, then both $T$ and $\mu$ may vary slowly in space and time, so that the relative variation in each mesoscopic cell is negligible. 

\vspace{.1cm}

\noindent The coherent part is given by a Hamiltonian and the observables evolve according to the Heisenberg equation. But, because of the local equilibrium approximation and because each mesoscopic cell is opened, the Hamiltonian will represent the {\em free energy} of the gas in the mesoscopic cell. Therefore it has the form $F=H-\mu N$ where $H$ represents the mechanical energy of the gas, while $N$ is the number of electrons (in the mesoscopic cell). The mechanical part $H$ of this Hamiltonian will be reduced to the potential energy of the electrons on the impurity sites, ignoring the hopping term between impurities, since the tunneling effect is so small for lightly doped semiconductors. The potential will be represented by a family 
$\epsilon=(\epsilon_x)_{x\in\Ll (s)}$ of independent identically distributed random variables, where $\epsilon_x$ belongs to the impurity band $\Delta$ which is a compact interval. For the sake of the present model, the distribution will be assumed to be uniform in $\Delta$
and it shall be independent of the random variables $s=(s_x)_{x\in\Ll}$.

\noindent The family of all random variables $s=(s_x)$ and ${\epsilon}=(\epsilon_x)$ 
will be simply denoted by a random variable $\omega=(s,\epsilon)$. The corresponding probability space consisting of the
compact set of all such families will be
denoted by $\Omega$ and the corresponding probability measure on the Borel sets of $\Omega$ by $\PM$.
Many operators such as the free energy depend on the randomness $\omega$, e.g. the free energy becomes
\begin{equation}
\label{mott09.eq-freeEner}
F(\omega)= \sum_{x\in\Ll (s)}
    (\epsilon_x-\mu)\;a_x^\dag\,a_x
\end{equation}
\noindent Since the electron-phonon interaction is weak, the dissipative part will be given in the Markov approximation, in which all degrees of freedom other than the electrons are integrated out, while the time scale for this interaction to rearrange the electron state is considered as negligible. Namely, the dissipative part of the dynamics will be described by the generator of a Markov semigroup, which, thanks to the Theorem by Lindblad \cite{Li76} is given by a Lindbladian of the form
\begin{equation}
\label{mott09.eq-lind}
\DG (A) = 
   \sum_{i} \left(
    \frac{1}{2}(L_i^\ast L_i A+AL_i^\ast L_i)- L_i^\ast A L_i
             \right)\,.
\end{equation}
\noindent The description of the model will precisely consist in proposing an expression for the $L_i$'s. In order to implement the Mott scheme, these operators should describe the jump of an electron form an impurity site $x\in\Ll(s)$ to an impurity site $y\in\Ll(s)$. 
The corresponding jump operator, denoted by $L_{x\to y}$,
should be proportional to $a_y^\dag a_x$, since the later annihilates an electron at $x$ and creates one at $y$. 
If one of the sites, $x$ or $y$, is not in the random lattice $\Ll(s)$, then there can't be a jump from $x$ to $y$. Hence we set $L_{x\to y}$ equal to zero in this case.
Hence,
\begin{equation}
\label{mott09.eq-Lxy}
L_{x\rightarrow y}(\omega) =
   s_x s_y\, \sqrt{\Gamma_{x\rightarrow y}}\;
     a_y^\dag a_x\,.
\end{equation}
\noindent where $\Gamma_{x\rightarrow y}$ is the probability rate at which the jump arises. Following the Mott argument, this jump probability rate needs to take into account the probability of a phonon being absorbed or created by the electrons. The absorption process is dominated by the probability of a phonon of energy $\epsilon$ to be created by the thermal bath. It is given by the Boltzmann factor $e^{-\beta \epsilon}$ with a good approximation\footnote{It should actually be given by the Bose-Einstein distribution $(e^{\beta\epsilon}-1)^{-1}$. But if $\beta\epsilon \gg 1$ it follows that $(e^{\beta\epsilon}-1)^{-1}\approx e^{-\beta\epsilon}$}. 
Then the electron located at $x$ absorbing this energy must find an available  site $y$ with energy $\epsilon_y=\epsilon+\epsilon_x >\epsilon_x$. On the other hand, if an electron at $x$ is in an excited state, it might decrease its energy by spontaneously emitting a phonon or energy $\epsilon\geq 0$, provided it finds a site $y$ at which the energy available is $\epsilon_y= \epsilon_x-\epsilon <\epsilon_x$. The spontaneous emission does not require any Boltzmann factor. Hence the probability for absorption and emission is proportional to $e^{-\beta (\epsilon_y-\epsilon_x)}$ if $\epsilon_y \geq\epsilon_x$ and to $1$ if $\epsilon_y \leq \epsilon_x$. Thus a unified formula for the probability of absorption or emission is $e^{-\beta (\epsilon_y - \epsilon_x)^+}$ where $\epsilon^+$ denotes the positive part of the real number $\epsilon$. 
This difference between emission and absorption leads to the quotient $\frac{\Gamma_{x\rightarrow y}}{\Gamma_{y\rightarrow x}}= e^{-\beta (\epsilon_y-\epsilon_x)}$ which is also known as {\em detailed balance} condition.

\vspace{.1cm}

\noindent In addition, in both cases, the electron must jump form $x$ to $y$ through a tunneling effect, namely decreasing the probability by a factor 
proportional to $e^{-|x-y|/r}$, if $r$ is the localization length. Note that in lightly doped semiconductors, $r$ is of the order of the Bohr radius $\rB$, since  the average distance between impurities corresponds approximately to $10 \rB$. This proportionality factor will be normalized by dividing it by $Z$, where
\begin{equation}
\label{mott09.eq-norm}
Z = \sum_{m\in\Ll (s)} e^{-|m|/r}\,.
\end{equation}
\noindent This leads to the expression of the jump rate proposed by Mott
\begin{equation}
\label{mott09.eq-Gammxy}
\Gamma_{x\rightarrow y}(\omega) = \Gamma_0 \;
   \frac{e^{-|x-y|/r}}{Z} \; 
    e^{-\beta(\epsilon_y-\epsilon_x)^+} \,,
\qquad
x,y\in\Ll (s)
\end{equation}
\noindent which gives the detailed balance condition
\begin{equation}
\label{mott09.eq-Gxy}
\frac{\Gamma_{x\rightarrow y}}{\Gamma_{y\rightarrow x}}=
  e^{-\beta (\epsilon_y-\epsilon_x)}\, .
\end{equation}

\begin{rem}
\label{mott09.rem-Gamma0}
In the present paper we assume that $\Gamma_0$ is a constant. The present results however remain valid if $\Gamma_0$ is a function of $\epsilon_x$, $\epsilon_y$ and $\beta$ satisfying the following properties: $\Gamma_0(\epsilon_x, \epsilon_y, \beta) = \Gamma_0(\epsilon_y,\epsilon_x, \beta)$, (hence the detailed balance condition (\ref{mott09.eq-Gxy}) will remain valid), and $\inf\{ \Gamma_0(\epsilon_x, \epsilon_y, \beta ): \epsilon_x, \epsilon_y \in \Delta , \beta \in (0, \infty ) \} >0$.
\end{rem}

\noindent The factor $\Gamma_0$ is a parameter with the dimension of the inverse of a time, fixing the order of magnitude of the effect. 
We let $\DG_\omega^{kin}$ denote the Lindblad operator obtained from this family of jump operators,
\begin{equation}
\label{mott09.eq-dgkin}
\DG_\omega^{kin}(A) =
  \sum_{x\neq y\in\Ll (s)}
   \left(\frac{1}{2}
    \left(
   L_{x\to y}^\ast L_{x\to y} A+AL_{x\to y}^\ast L_{x\to y}
    \right) -
     L_{x\to y}^\ast A L_{x\to y}\right) \;.
\end{equation}
However, this part of the model is insufficient to describe the return to equilibrium, because $\DG^{kin}$ leaves the number operator $N$ invariant, as can be checked easily. Therefore it does not take into account the possibility for an electron to jump out or to jump in the system (thermal bath). An electron can jump out of the system in two ways: (i) either its energy becomes too large or too small to stay within the impurity band $\Delta$, or (ii) the electron is kicked out of the mesoscopic cell under consideration. Similarly the opposite processes arises to allow an electron to jump in the system. A standard way to take this process into account is to create an extra site $\star$ called the {\em cemetery} and to describe these processes as a simple jump ${\star\rightarrow x}$ or ${x\rightarrow \star}$. In the first case, an electron is created at $x$, while in the other it is annihilated at $x$. The cemetery really describes the thermal bath and it is only natural to interpret the chemical potential $\mu$ as the energy associated with this new site. Hence, the corresponding jump operators will be given by
\begin{equation}
\label{mott09.eq-Lxstar}
L_{x\rightarrow \star}(\omega) =
   s_x\left(\Gamma_{x\rightarrow \star}\right)^{1/2} a_x\,,
\quad
   L_{\star\rightarrow x}(\omega)=
   s_x\left(\Gamma_{\star\rightarrow x}\right)^{1/2} a_x^\dag\,.
\end{equation}
\noindent A convenient way to define the jump rates $\Gamma_{x\rightarrow \star}$ is to imitate what was done earlier and to consider the sites $\star$ similar as the sites occupied by electrons, associated with the energy $\mu$, leading to
\begin{equation}
\label{mott09.eq-Gammxstar}
\Gamma_{x\rightarrow \star}(\omega ) = \Gamma_\star \;
  e^{-\beta(\mu-\epsilon_x )^+}\,, \quad
   \Gamma_{\star\rightarrow x}(\omega)  = \Gamma_\star \;
    e^{-\beta(\epsilon_x-\mu )^+} \,,
\quad
    \frac{\Gamma_{x\rightarrow \star}}
           {\Gamma_{\star\rightarrow x}}=
     e^{\beta(\epsilon_x -\mu)}\;
\end{equation}
where $x\in\Ll(s)$.
\noindent Here again, $\Gamma_\star >0$ is a jump rate probability fixing the time scale for the cemetery process.

\begin{rem}
\label{mott09.rem-Gamma2}
In the present paper  $\Gamma_\star$ is a constant. The present results however remain valid if $\Gamma_\star$ is a function of $\epsilon_x$,  and $\beta$ satisfying $\inf\{ \Gamma_\star(\epsilon_x, \beta ): \epsilon_x  \in \Delta , \beta \in (0, \infty ) \} >0$. See also the related Remarks~\ref{mott09.rem-Gamma0} and \ref{mott09.rem-Gamma3}.
\end{rem}

\noindent The new jump operators describing the exchange between the thermal bath and the electron gas gives rise to a Lindblad operator denoted by $\DG_\omega^\star$. However, due to the anti-commutation rules, the Lindbad operator, acting on an observable $A$, takes the form
\begin{eqnarray}
\label{mott09.eq-dgstar}
\DG_\omega^\star(A) &=& 
   \sum_{x\in\Ll (s)} \left(\frac{1}{2}
    \left(
   L_{x\to\star}^\ast L_{x\to\star} A+AL_{x\to\star}^\ast L_{x\to\star}
    \right) - (-1)^{d_A}L_{x\to\star}^\ast A L_{x\to\star}\right) +\nonumber\\
&&  
    \sum_{x\in\Ll (s)} \left(\frac{1}{2}
    \left(
   L_{\star\to x}^\ast L_{\star\to x} A+AL_{\star\to x}^\ast L_{\star\to x}
    \right) - (-1)^{d_A} L_{\star\to x}^\ast A L_{\star\to x}\right)
\end{eqnarray}
\noindent where $d_A$ is the degree of $A$ given by the natural $\ZM_2$ grading of the CAR algebra $\AG(s)$. More details will be given below.
Hence, the total dynamics is described by an operator $\LG$ acting on the set of observables by
\begin{equation}
\label{mott09.eq-LindMod}
\LG_\omega(A) \,=\, \imath [F(\omega),A] \,+\, \DG_\omega^{kin}(A) \,+\, \DG_\omega^\star(A)\,.
\end{equation}
\noindent What is left for the mathematician, is to make sure that this description does not produce any hidden effect that could be related with the infinite volume limit. For indeed, the mesoscopic cells have an undefined size, only the order of magnitude of the size is fixed. In addition, the random character of the set $\Ll (s) $ of the impurity sites and of the $\epsilon_x$'s must be included in the description to make sure that the results obtained are almost surely independent of the configuration of the impurities. These two requirements are the very reason why the mathematical formalism is so demanding. 


\section{The Coherent Evolution}
\label{mott09.sect-coherent}

\noindent In this section the coherent evolution of the quantum motion
is studied and the equilibrium state of the unperturbed system is obtained.
The coherent evolution is a one parameter group automorphism on the algebra
of the observables.  


 \subsection{Observables}
 \label{mott09.ssect-obs}

\noindent In this subsection the CAR algebra of the observables is studied.
Recall that a CAR algebra is a \Cs generated by a countable number of creation and annihilation operators which satisfy the rules stated in eq.~(\ref{mott09.eq-car}). It is well-known, and elementary to show, that the complex algebra generated by the pair $a_x,a_x^\dag$ is isomorphic 
to the set $M_2(\CM)$ of $2\times 2$ matrices by using the analogy
$$
a_x \rightarrow \sigma^- =
\left[
\begin{array}{cc}
0 & 0\\
1 & 0
\end{array}\right]\,,
\qquad
a_x^\dag \rightarrow \sigma^+ = 
\left[
\begin{array}{cc}
0 & 1\\
0 & 0
\end{array}\right]\,.
$$
\noindent If a finite family $\Lambda\subset \Ll(s)$ of sites is considered instead, the $\ZM_2$-graded algebra generated by the family $\{a_x,a_x^\dag\,;\, x\in \Lambda\}$ is isomorphic to the tensor product $M_2(\CM)^{\otimes \Lambda}$ but the previous analogy ought to be modified in order to insure that $a_x$ and $a_y$ anti-commute. One way to describe such an isomorphism 
is given by the so-called {\em Jordan-Wigner transformation}: since $\Lambda$ is finite let its elements be numbered from $1,2,\cdots, m=|\Lambda|$. Then 
\begin{equation}
\label{mott09.eq-jw}
a_k \rightarrow 
   \underbrace{\sigma_3\otimes \cdots \sigma_3}_{k-1}\otimes 
    \sigma^-\otimes 
     \underbrace{\id_2\otimes \cdots \otimes \id_2}_{m-k}\,,
\quad
a_k^\dag \rightarrow 
   \underbrace{\sigma_3\otimes \cdots \sigma_3}_{k-1}\otimes 
    \sigma^+\otimes 
     \underbrace{\id_2\otimes \cdots \otimes \id_2}_{m-k}\,,
\end{equation}
\noindent where 
$$
\sigma_3 = 
\left[
\begin{array}{cc}
1 & 0\\
0 & -1
\end{array}\right]\,.
$$
\noindent As can be seen, the Jordan-Wigner transformation is not canonical in that it requires the choice of an order on $\Lambda$. 
However it can sometimes be convenient for practical uses.

\vspace{.1cm}

\noindent A {\em canonical} description of the observable algebra has been given in the past and the main references are \cite{BrRo1,BrRo2}. 
Let this construction be summarized here. As shown before, there is a non-canonical isomorphism between $\AG_\Lambda(s)$ and 
$M_2(\CM)^{\otimes \Lambda\cap\Ll(s)}$, so that $\AG_\Lambda(s)$ has dimension $2^{|\Lambda\cap\Ll(s)|}$. If $\Lambda \subset \Lambda'$ 
there is a canonical isometric embedding $i_{\Lambda,\Lambda'}:\AG_{\Lambda}(s)\mapsto \AG_{\Lambda'}(s)$ (in the Jordan-Wigner 
representation (\ref{mott09.eq-jw}) this embedding consists simply in adding to the $a_k$'s one more factor $\otimes \id_2$ at 
the end of the chain). The algebras $\AG_\Lambda(s)$ are called {\em local}, while the inductive limit
\begin{equation}
\label{mott09.eq-qlocal}
\AG(s) = \lim_{\rightarrow} 
   \left(\AG_\Lambda(s),i_{\Lambda,\Lambda'}\right)\,,
\end{equation}
\noindent is equal to the {\em quasi-local} observable algebra.
It should be remarked at this point that $\AG(s)$ is {\em random} as it depends upon the random variable $s=(s_x)_{x\in\Ll}$. 
However, since $s\in \{0,1\}^\Ll =\Xi$ and since $\Xi$ can be considered as a compact space, this family of algebras can be seen as a 
{\em continuous field} of \Css \cite{Dix,Ty58,Ty62,Ty63}.

\noindent Recall that a \Cs algebra $\AG$ is $\ZM_2$ graded when there exists a $\star$-automorphism $\sigma$
on $\AG$ which satisfies $\sigma^2= \id_{\AG}$, (in other words, when the group $\ZM_2$ acts on $\AG$). The 
$\star$-automorphism $\sigma$ is called grading. In the case of the CAR algebra, the grading is the canonical
$\star$-automorphism defined by $\sigma (a_x) = - a_x$ for all $x$. If $\AG$ is a $\ZM_2$ graded \Cs with grading 
$\sigma$ then the elements $A \in \AG$ with $\sigma (A)= A$ are called even, and the elements $A \in \AG$ 
with $\sigma (A)= -A$ are called  odd. The degree of every even element $A$ is defined to be equal to zero,
(denoted by $d_A=0$), and the degree of every odd element $A$ is defined to be equal to one, (denoted by 
$d_A=1$). In the above Jordan-Wigner representation, the operators of degree zero are represented by diagonal matrices, while the operators of degree one are given by the off diagonal ones.  
\begin{proposition}
\label{mott09.prop-contf}
The family $\{\AG(s)\,;\, s\in \Xi=\{0,1\}^\Ll \}$ can be endowed with the structure of a continuous field of $\ZM_2$-graded \CsS. 
\end{proposition}
\noindent {\bf Proof: } Recall first \cite{Dix} that a section of the field $\left(\AG(s)\right)_{s\in\Xi}$ is an element of the
 Cartesian product $\prod_{s\in\Xi}\AG(s)$. To define a structure of a continuous field it is necessary to define first a set $\fs$ of
 {\em reference sections} with the property that (i) for each $s\in\Xi$ the set $\{\xi(s)\,;\, \xi\in \fs\}$ is dense in $\AG(s)$ and 
(ii) if $\xi\in\fs$, the map $s\in\Xi\mapsto \|\xi(s)\|\in \RM_+$ is continuous. Then a {\em continuous section} is an element 
$\eta \in \prod_{s\in\Xi}\AG(s)$ such that for any $s\in\Xi$ and any $\epsilon >0$ there is a neighborhood $\us$ of $s$ in $\Xi$ 
and a reference section $\xi\in \fs$ such that $\|\xi(s')-\eta(s')\| <\epsilon$ for $s'\in\us$. 

\vspace{.1cm}

\noindent In the present situation it ought to be remarked that the product topology on $\Xi$ is defined through the set of open sets
 $\us_\Lambda(s) = \{s'\in \Xi\,;\, s_x'=s_x\,, \forall x\in \Lambda \}$. Hence $s'$ is close to $s$ if there is some finite subset 
$\Lambda\subset \Ll$ such that $s_x'=s_x$ for $x\in\Lambda$. Then let $a_x$ be the section defined by $a_x(s)=s_x a_x \in \AG(s)$. 
Let $\fs$ be the $\ast$-algebra defined by the $a_x$'s. Namely $\xi\in\fs$ if and only if there is $\Lambda \subset \Ll$ finite, 
an integer $M\in\NM$ and continuous functions $\{\lambda_{x_1,\ldots, x_m}^{\sharp_1,\ldots \sharp_m}\in \Cc(\Xi)\}$ such that
\begin{equation}
\label{mott09.eq-secfs}
\xi(s) = \sum_{m=0}^M 
   \sum_{x_1,\ldots, x_m \in \Lambda}
    \sum_{\sharp_1,\ldots \sharp_m}
     \lambda_{x_1,\ldots, x_m}^{\sharp_1,\ldots \sharp_m}(s) \;
      s_{x_1}\cdots s_{x_m} \;
       a_{x_1}^{\sharp_1}\cdots a_{x_m}^{\sharp_m}
\end{equation}
\noindent where the $\sharp_i$'s $\in \{.,\dag\}$ label the $a_x, a_x^\dag$'s. Clearly, if $\Lambda\subset \Lambda'$ and if
 $s'\in \us_{\Lambda'} (s)$ the previous expression for $\xi(s)$ does not change as $s'$ replace $s$, so that 
$s\in\Xi \mapsto \|\xi(s)\|\in \RM_+$ is continuous. Moreover, the set $\{\xi(s)\,;\, \xi\in\fs \}$ generate $\AG(s)$ as a \CS. 
Hence $\fs$ satisfy the conditions.
\hfill $\Box$

\begin{proposition}
\label{mott09.prop-cov}
The continuous field $\left((\AG(s)_{s\in\Xi}, \fs\right)$ is covariant by the translation group, namely, for any
 $a\in\ZM^d$, there is a $\ZM_2$-graded $\ast$-isomorphism $\eta_a :\AG(s)\mapsto \AG(\tra^as)$ leaving $\fs$ invariant.
\end{proposition}

\noindent {\bf Proof: } It is elementary to check that the $\ast$-isomorphism $\eta_a$ is defined by 
$\eta_a(a_x) (s)= s_{x} a_{x+a}$ exists and does the job since $(\tra^a s)_{x+a}= s_x$. Moreover it is easy to check that
 $\eta_a\circ \eta_b = \eta_{a+b}$, that $\eta_0=id$ and that $\eta_a$ commutes with the grading. 
\hfill $\Box$

\vspace{.2cm}

\noindent The continuous field $\left((\AG(s))_{s\in\Xi}, \fs\right)$ contains an important sub-field made of commutative \CsS. 
For given $s\in\Xi$, let $\CG(s)$ be the closed sub-algebra of $\AG(s)$ generated by the family $\{s_x n_x\,;\, x\in \Ll\}$ where
$n_x=a_x^\dag a_x$. It is
easy to see that $\CG(s)$ is commutative, it contains only elements of degree zero, it is generated by projections and generates also
a continuous $\ZM^d$-covariant field of \CsS.

\vspace{.1cm}

\noindent Two new ingredients will be needed in the rest of the paper. First it will be convenient to consider the field 
$\AG$ as a field over the compact space $\Omega$ instead. The only difference will be that, in the eq.~(\ref{mott09.eq-secfs}), 
the coefficients will be allowed to depend continuously upon $\omega= (s,\epsilon)$. 
When necessary this field will be denoted by 
$\AG(\omega)$, and similarly the corresponding commutative sub-field will be denoted by $\CG(\omega)$. The following can be found in \cite{Dix}.

\begin{corollary}
\label{mott09.cor-contfiO}
The field $\AG$ can be continued as a continuous field $\AG= \left(\AG(\omega) \right)_{\omega\in\Omega}$ of \Css which is covariant 
with respect to the translation group. Similarly $\CG=\left(\CG(\omega) \right)_{\omega\in\Omega}$ will denote the abelian sub-field extending
 $\left(\CG(s)\right)_{s\in\Xi}$.  
\end{corollary}

\noindent The other ingredient is the use of {\em groupoids} to describe the $\ZM^d$ action \cite{Co78,Re80}. Here the group $\ZM^d$ acts on the 
space $\Omega$ by 
$\tra^a(\omega)=\tra^a(s,\epsilon)= (\tra^a(s),\tra^a(\epsilon))$
where $a\in\ZM^d$, $(\tra^a(s))_x=s_{x-a}$ and $(\tra^a(\epsilon))_x=\epsilon_{x-a}$.
Consequently, the {\em crossed product} 
$\Gamma_\Omega= \Omega\rtimes \ZM^d$ is a groupoid described as follows: (i) elements are pairs $\gamma= (\omega,a) \in \Omega\times \ZM^d$, 
(ii) $\Omega$ is the set of {\em objects} or of {\em units}, (iii) each element has a {\em range} and a {\em source} in $\Omega$, 
here $r(\omega,a)=\omega\,,\, s(\omega,a)=\tra^{-a}\omega$, (iv) two elements $\gamma$ and $\gamma'$ are composable if $s(\gamma)=r(\gamma')$, 
namely if $\gamma=(\omega,a)$ then $\gamma' =(\tra^{-a}\omega,a')$ and there is a composition law $\gamma\circ\gamma'$ which is here given
 by $(\omega,a+a')$, (v) the elements of the form $(\omega,0)$ are units and can be identified with the points in $\Omega$, (vi) each element
 $\gamma$ admits an inverse with exchange of range and source, namely here $\gamma^{-1}= (\tra^{-a}\omega,-a)$. The groupoid
 $\Gamma_\Omega$ will be endowed with the product topology and it is elementary to check that all groupoid maps defined above are continuous. 

\vspace{.1cm}

\noindent To express the covariance of the field $\AG= \left(\AG(\omega) \right)_{\omega\in\Omega}$, it is convenient to see the translation 
$\eta_a$ as a function of the groupoid variables instead, namely if $\gamma =(\omega,a)$, then $\gamma$ can be seen as sending its 
source $s(\gamma)= \tra^{-a}\omega$ into its range $r(\gamma)=\omega$, so that 
$$
\eta_{(\omega,a)}: \AG(\tra^{-a}\omega) \mapsto \AG(\omega)\;.
$$

\begin{definition}
\label{mott09.def-cov}
A field $\left(\theta_{\omega}\right)_{\omega\in\Omega}$ of $\ast$-homomorphisms from $\AG(\omega)$ into itself will be called 
{\em covariant} if the following diagram is commutative
\begin{equation*}
\begin{array}{ccc}
   \AG(\tra^{-a}\omega) & 
    \stackrel{\eta_{(\omega,a)}}{\longrightarrow} & 
     \AG(\omega)\\
      \theta_{\tri^{-a}\omega} \downarrow & &
       \theta_\omega \downarrow\\
        \AG(\tra^{-a}\omega) & 
         \stackrel{\eta_{(\omega,a)}}{\longrightarrow} & 
          \AG(\omega)
       \end{array}\;.
\end{equation*}
\noindent This field will be called continuous whenever it transforms every continuous section of $\AG(\omega)$ into a continuous section.
\end{definition}


 \subsection{The coherent dynamics}
 \label{mott09.ssect-dyn}

\noindent Recall that the coherent part of the quantum motion is a group automorphism acting on the CAR algebra $\AG$. Let $\Lambda \subset \Ll$  be a finite set. Then the particle number $N_\Lambda$, the Hamiltonian $H_\Lambda$ and the free energy $F_\Lambda$ within $\Lambda$ are defined as follows:
\begin{equation}
\label{mott09.eq-nbodyNH}
N_\Lambda(\omega) =
    \sum_{x\in \Lambda} s_x \;n_x\,,
\hspace{1cm}
H_\Lambda (\omega) = 
  \sum_{x\in \Lambda} s_x \epsilon_x\;n_x\,,
\end{equation}
\begin{equation}
 \label{eq-FreeEnergy}
F_\Lambda (\omega )= 
 H_\Lambda (\omega )- \mu N_\Lambda (\omega )= 
  \sum_{x \in \Lambda } s_x (\epsilon_x - \mu) n_x \,,
\end{equation}
where
\begin{equation*}
 n_x=a_x^\dag a_x\quad \text{and}\quad \omega= (s,\epsilon)\in \Omega\,.
\end{equation*}

\noindent The finite volume coherent dynamics is generated by the free energy operator
\begin{equation}
\alpha_t^{(\omega,\Lambda)} (A) = 
   e^{\imath t(H_\Lambda -\mu N_\Lambda)}
    \;A\;
     e^{-\imath t (H_\Lambda -\mu N_\Lambda)}\,,
\qquad
A\in \AG_\Lambda( \omega )\,,\; t\in \RM\,.
\end{equation}
\noindent Usual arguments \cite{BrRo2}, which become trivial in the present situation, show that the infinite volume limit exists, namely
\begin{equation}
\alpha_t^{(\omega)}(A) = 
   \lim_{\Lambda \uparrow \Ll} \alpha_t^{(\omega,\Lambda)} (A)\,,
\qquad
A\in \AG_{\Lambda_0}(s)\,, t\in \RM\,.
\end{equation}
\noindent Indeed, it is elementary to show that
\begin{equation}
\label{mott09.eq-dyns}
\alpha_t^{(\omega)}(s_xa_x^\dag) = 
   e^{\imath ts_x(\epsilon_x -\mu)}s_xa_x^\dag\,,
\qquad
  \alpha_t^{(\omega)}(s_xa_x) = 
   e^{-\imath ts_x(\epsilon_x -\mu)}s_xa_x\,,
\end{equation}
\noindent since it is already true for $\alpha_t^{(\omega,\Lambda)}$ as soon as $\Lambda \ni x$. Hence, $\alpha_t^{(\omega)}$ can be computed 
on any monomial, thus on any polynomial, in the creation-annihilation operator. Therefore $\alpha_t^{(\omega)}$ is defined anywhere in $\AG(s)$. 
It also follows from these formulas that $s_xn_x$ is invariant by the dynamics, a fact which comes from ignoring the hopping terms due to possible tunneling between impurity sites. As a consequence, the elements of $\CG$ are left invariant by the dynamics. To summarize, thanks to the 
Definition~\ref{mott09.def-cov}, the following holds (the proof is left to the reader).

\begin{proposition}
\label{mott09.prop-dyn}
The field of equilibrium dynamics $\alpha^{(\omega)}= \left(\alpha_t^{(\omega)}\right)_{t\in\RM}$ induced by the Hamiltonian (\ref{mott09.eq-nbodyNH}) on each $\AG(\omega)$ is well defined, $\ZM^d$-covariant and continuous. Moreover its restriction to the sub-field $\CG$ is trivial. 
\end{proposition}

\noindent The generator of the coherent evolution $\alpha^{(\omega)}_t$ is given by the commutator with the free energy operator defined in eq.~(\ref{eq-FreeEnergy}), {\em i.e.}
\begin{equation} 
\delta_\omega (A) = 
    \frac{d\alpha^{(\omega)}_t(A)}{dt}\bigg|_{t=0}
\;= \lim_{\Lambda \uparrow \Ll} \imath [ F_\Lambda (\omega ) , A ] \quad \mbox{ for } \quad
A \in \AG_{loc} \, .
\end{equation}

 \subsection{Equilibrium State}
 \label{mott09.ssect-eq}

\noindent Once the dynamics is defined, the next step is to find the possible equilibrium states. There are two ways to do that. The first is to consider 
the finite volume approximations and establish the quantum version
of the DLR equations. The other one, valid in the infinite volume limit, consists in implementing the Kubo-Martin-Schwinger conditions (KMS) which were introduced in \cite{K,MS} and further studied in \cite{HHW,Ta70,BrRo2,Ta03}. 
In the present situation, due to the extreme simplicity of the dynamics, both approaches can be used and lead to the same explicit result. 
The KMS condition will be 
used here. 

\begin{definition}
\label{mott09.def-KMS}
Let $\rho$ be a state on a \Cs $\AG$, and let $\alpha= (\alpha_t)_{t\in \RM}$ be a one-parameter group of $\ast$-automorphisms of $\AG$. Then $\rho$ will 
be called $\beta$-KMS with respect to $\alpha$ if 

\noindent (i) it is invariant by the dynamics, namely $\rho\circ \alpha_t=\rho$ for all $t\in\RM$, 

\noindent (ii) if $A,B\in\AG$, then
\begin{equation}
\label{mott09.eq-KMS}
\quad\qquad\rho\left(AB\right) =
   \rho\left(\alpha_{-\imath \beta}(B)A\right)
\hfill \hspace{3cm}
 \mbox{\rm \bf (KMS condition)}
\end{equation}
\end{definition}

\noindent Note that (i) is a trivial consequence of (ii) if $\AG$ is unital, but it is convenient to define this condition in this way. In the 
present situation the following holds.
\begin{proposition}
\label{mott09.prop-gs}
For $\PM$-almost every $\omega= (s,\epsilon)\in\Omega$, there is a unique $\beta$-KMS state $\rho_\omega$ on $\AG(\omega)$ with respect to the dynamic 
$\alpha^{(\omega)}$. It is defined by 
\begin{eqnarray}
\label{mott09.eq-rhoprod}
\rho_\omega\left(\prod_{x\in \Lambda}A_x\right) &=& 
   \prod_{x\in \Lambda} \rho_\omega(A_x)\,,
\qquad
  A_x\in\AG_{\{x\}}(s)\\
\label{mott09.eq-rhoax}
\rho_\omega(s_xa_x^\sharp) &=& 0\,, \\
\label{mott09.eq-rhonx}
\rho_\omega(s_x n_x) &=& 
    s_x \frac{1}{1+e^{\beta(\epsilon_x-\mu)}}\,.
\end{eqnarray}
\noindent Moreover, the field of such states is continuous and covariant with respect to translations, namely 
$\rho_\omega\circ\eta_{\omega,a}= \rho_{\tri^{-a}\omega}$. 
\end{proposition}

\noindent {\bf Proof: } (i) In order to prove eq.~(\ref{mott09.eq-rhoax}) and (\ref{mott09.eq-rhonx}), it is sufficient to 
assume that $s_x=1$. Then, each element of the elementary algebra $\AG_{\{x\}}$ can be written as a linear combination of 
$\id, a_x, a_x^\dag$ and $n_x= a_x^\dag a_x$. Thanks to eq.~(\ref{mott09.eq-dyns}) and since a KMS-state is time invariant, 
it follows that $\rho_\omega (a_x) = \rho_\omega (\alpha_t(a_x))= e^{-\imath t (\epsilon_x-\mu)} \rho_\omega (a_x)$ for all 
$t\in\RM$. Since the distribution of $\epsilon_x$ is absolutely continuous, $\epsilon_x\neq \mu$ $\PM$-almost surely. Hence 
$\rho_\omega (a_x) = 0$. In much the same way $\rho_\omega (a_x^\dag) = 0$. Similarly, if $A, B\in \AG_\Lambda(\omega)$ where 
$x\notin \Lambda$, the same argument shows that $\rho_\omega (A a_x B) = \rho_\omega (A a_x^\dag B)= 0$. Consequently if 
$A\in \AG(\omega)\setminus \CG(\omega)$ it follows by induction on $\Lambda$ and by density, that $\rho_\omega (A)= 0$. Hence it is 
sufficient to reduce the analysis on the commutative sub-algebra $\CG(\omega)$. 

\vspace{.1cm}

\noindent (ii) Since $n_x$ commutes to any element of $\CG_\Lambda(\omega)$, it is sufficient to consider an expression of the form 
$\rho_\omega (An_x) = \rho_\omega (Aa_x^\dag a_x)$ where $A$ is a local observable in $\CG(\omega)$ with support not meeting $x$. Using the 
KMS-condition, this gives
$$
\rho_\omega (An_x) = \rho_\omega (\alpha_{-\imath\beta}(a_x) Aa_x^\dag) = e^{-\beta(\epsilon_x-\mu)} \rho_\omega (a_xAa_x^\dag)\,.
$$ 
Since the support of $A$ does not meet $x$ and $A\in\CG(\omega)$, it follows that $A$ commutes with $a_x$ so that
 $\rho_\omega (a_x A a_x^\dag) = \rho_\omega (A a_xa_x^\dag) = \rho_\omega (A(\id -n_x))$. Hence this gives
$$\rho_\omega(A n_x) = 
   \frac{\rho_\omega (A)}{1+e^{\beta(\epsilon_x-\mu)}}
$$
\noindent Eliminating points of the support of $A$ one after another leads to the formulas above. 

\vspace{.2cm}

\noindent (iii) It is obvious that the field $\rho= \left(\rho_\omega\right)_{\omega\in\Omega}$ of states is continuous. For indeed it is
 enough to show that the map $\omega\in\Omega\mapsto \rho_\omega(\xi(\omega))\in\CM$ is continuous for any continuous section of the field $\AG$. 
By definition of continuous sections, it is sufficient to chose $\xi\in\fs$. But this is exactly choosing $\xi(\omega)$ as a polynomial in
 the creation-annihilation operators. The formulas~(\ref{mott09.eq-rhoprod}), (\ref{mott09.eq-rhoax}) and (\ref{mott09.eq-rhonx}) show immediately the 
continuity with respect to $\omega$. 

\vspace{.1cm}

\noindent (iv) Similarly, the covariance with respect to translation follows immediately from the 
formulas~(\ref{mott09.eq-rhoprod}), (\ref{mott09.eq-rhoax}) and (\ref{mott09.eq-rhonx}).
\hfill $\Box$

\begin{proposition}
\label{mott09.prop-faithful}
For all $\omega\in\Omega$ the state $\rho_\omega$ is faithful.
\end{proposition}

\noindent {\bf Proof: } First the state $\rho_\omega$ is defined by the 
formulas~(\ref{mott09.eq-rhoprod}), (\ref{mott09.eq-rhoax}) and (\ref{mott09.eq-rhonx}).
 It is easy to check that such result can also be obtained from (omitting the reference to $\omega$)
\begin{equation*}
\rho(A) = 
   \frac{\TR\left(e^{-\beta F_\Lambda}A\right)}{\Zz(\Lambda)}\,,
\qquad
  A\in\AG_\Lambda\,,
\end{equation*}
\noindent where $\Zz(\Lambda)$ is a normalization constant and $F_\Lambda = H_\Lambda-\mu N_\Lambda$. In particular, the restriction of 
$\rho$ to $\AG_\Lambda$ is faithful. Moreover, there is a conditional expectation $\EM_{\Lambda}:\AG \mapsto \AG_\Lambda$ defined  by
\begin{equation*}
\EM_{\Lambda}(A) = 
   \EM_{\Lambda',\Lambda}(A) =
     \frac{\TR_{\Lambda'\setminus\Lambda}
      \left(e^{-\beta F_{\Lambda'\setminus\Lambda}}A\right)}
       {\Zz(\Lambda)}\,,
\qquad
\mbox{\rm if}\;\; A\in\AG_{\Lambda'}\,.
\end{equation*}
\noindent It is elementary to check that $A\geq 0 \Rightarrow \EM_\Lambda(A) \geq 0$, that $\EM_\Lambda(\id) =\id$, so that 
$\| \EM_\Lambda(A)\|\leq \|A\|$ for $A\in \AG$. Moreover, it is easy to check that $\rho\circ \EM_\Lambda = \rho$ for all finite $\Lambda\subset \Ll$. 

\vspace{.1cm}

\noindent Let now $A\in\AG$ be positive and such that $\rho(A)=0$. Then $\rho(A) = \rho(\EM_\Lambda(A))=0$. Since $\rho$ is faithful 
on $\AG_\Lambda$ it follows that $\EM_\Lambda(A)=0$ for all finite $\Lambda$'s. 
Let $\varepsilon>0$, there is $\Lambda_\varepsilon\subset \Ll$
 finite and $A_\varepsilon \in\AG_{\Lambda_\varepsilon}$ such that $\| A-A_\varepsilon\|<\varepsilon/2$. Hence, whenever 
$\Lambda \supset \Lambda_\varepsilon$, $\| \EM_\Lambda(A-A_\varepsilon) \| = \|A_\varepsilon\| <\varepsilon/2$. Therefore 
$\|A\|\leq \varepsilon$ for any $\varepsilon$ implying $A=0$. 
\hfill $\Box$


 \subsection{The GNS Representation}
 \label{mott09.ssect-gns}

\noindent The Gelfand-Na\u{\i}mark-Segal construction (GNS) is a fundamental tool in the study of \CsS. 
Let $\AG$ be a \CS, for convenience it will be assumed to be unital. Let $\rho$ be a state on $\AG$. Then a Hilbertian inner product can be 
defined through $\langle A| B\rangle = \rho(A^\ast B)$ for $A,B\in \AG$. The set $\NNG$ of elements $A\in \AG$ for which $\rho(A^\ast A)=0$ 
is a closed left $\AG$-module. The quotient space $\AG/\NNG$ inherits the structure of a separated pre-Hilbert space. By completion it gives a 
Hilbert space $\Hh=L^2(\AG, \rho)$ and a canonical linear map $\zeta: A\in\AG \mapsto \zeta(A)\in \Hh$ such that
\begin{equation}
\langle \zeta(A) | \zeta(B)\rangle =
   \rho(A^\ast B)\,.
\end{equation}
\noindent If $\AG$ is abelian, then, by Gelfand's theorem, it is isomorphic to the set of continuous functions on some compact space $X$, unique up 
to homeomorphism, called the spectrum of $\AG$. Then a state is just a probability measure on $X$ and $\Hh=L^2(X,\rho)$. 

\vspace{.1cm}

\noindent $\Hh$ inherits the structure of a left $\AG$-module so that the map $\pi(A): \zeta(B)\in \Hh \mapsto \zeta(AB)\in\Hh$ is well defined and
 extends to a representation of $\AG$ in $\Hh$. The vector $\xi= \zeta(\id)$ is cyclic since $\zeta(A)=\pi(A)\xi$ so that $\Hh$ is the closure of 
the set of $\pi(A)\xi$ as $A$ varies in $\AG$. The weak closure of $\pi(\AG)$ is a von Neumann algebra denoted by $L^{\infty}(\AG,\rho)= \MG$. Clearly,
$\rho$ extends as a normal state on $\MG$ since
\begin{equation}
\rho(A) = \langle \xi |\pi(A) \xi\rangle\,.
\end{equation}
\noindent If $\AG$ is abelian with spectrum $X$, the representation is given by point wise multiplication in $L^2(X,\rho)$. The vector $\xi$ is the
 constant function equal to one and the von Neumann algebra is the space $L^\infty(X,\rho)$ of essentially bounded $\rho$-measurable functions on $X$. 

\vspace{.1cm}

\noindent If $\alpha$ is a $\ast$-automorphism of $\AG$ leaving $\rho$ invariant, it defines a unitary operator 
$U_\alpha:\zeta(A)\mapsto \zeta(\alpha(A))\in \Hh$ such that
\begin{equation}
U_\alpha \pi(A) U_\alpha^{-1} = 
   \pi(\alpha(A))\,,
\quad
A\in \AG\,,
\qquad
    U_\alpha \xi = \xi\,.
\end{equation}

\vspace{.1cm}

\noindent The Tomita-Takesaki theory \cite{Ta70} is based upon the densely defined conjugate linear operator $S$ defined by $S\zeta(A) = \zeta(A^\ast)$. The main result of Tomita and Takesaki is that $S$ is well defined, closable and that, if $S$ denotes also the closure, $\Delta= S^\ast S$ is a positive self-adjoint operator on $\Hh$ called the {\em modular operator}. The polar decomposition $S= J\Delta^{1/2}$, defines a conjugate linear involution $J$ such that $J\MG J$ coincides with the commutant of $\MG$ on $\Hh$. It has been shown \cite{Ta70} that there is a group of $\ast$-automorphism $\theta$ on $\MG$ satisfying a KMS-condition, namely
\begin{equation*}
\theta_t(M) = \Delta^{\imath t} M \Delta^{-\imath t}\,,
\qquad
 \rho(M_1M_2) = \rho(\theta_{-\imath}(M_2) M_1)\,,
\qquad
 M, M_1, M_2\in \MG\,.
\end{equation*}
\noindent In the commutative case the modular operator is trivial while $JA$ coincides with the complex conjugate of $A$. In the 
early seventies, Araki \cite{Ar74} and Connes \cite{Co74} realized that the set of $\zeta(A)$ with $A\in\AG, A\geq 0$ generates a cone $\Hh_+$ defined by
\begin{equation*}
\Hh_+ = \overline{
    \{\Delta^{1/4}\pi(A)\xi\,;\, A\in\AG\,,\, A\geq 0\}
           }\,,
\qquad
   J\Hh_+ =\Hh_+\,.
\end{equation*}
\noindent Connes characterized such a positive cone in $L^2(\AG,\rho)$ as being {\em self-dual, homogeneous} and {\em oriented}.
 It is the non-commutative analog of the set of positive square integrable functions.

\vspace{.2cm}

\noindent In the present situation the general theory gives the following result.

\begin{proposition}
\label{mott09.prop-GNS}
Let $\AG= \big(\AG(\omega)\big)_{\omega\in \Omega}$ be the continuous field of \Css defined in Section~\ref{mott09.ssect-obs}. Let 
$\alpha= \big(\alpha^{(\omega)}\big)_{\omega\in \Omega}$ be the field of dynamics defined in Section~\ref{mott09.ssect-dyn}. Then the 
GNS-construction leads to a continuous field $\Hh= (\Hh_\omega)_{\omega\in \Omega}$ of Hilbert spaces, a continuous section 
$\xi=(\xi_\omega)_{\omega\in \Omega}$ of unit vectors, with a $\ZM^d$ action, namely a unitary representation of the groupoid 
$\Gamma_\Omega$, given by unitary maps $V_{\omega,a}:\Hh_{\tri^{-a}\omega}\mapsto \Hh_{\omega}$ satisfying
$$V_{\omega, a+b} = V_{\omega, a}V_{\tri^{-a}\omega, b}\,,
\qquad
 V_{\omega, a}\xi_{\tri^{-a}\omega} = \xi_{\omega}\,.
$$

\noindent The $\ZM_2$-grading in $\AG$ is represented by a continuous covariant field $G=(G_\omega)_{\omega\in \Omega}$ of operators 
satisfying $G_\omega=G_\omega^\ast= G_\omega^{-1}$ and the covariance condition
$$V_{\omega,a}G_{\tri^{-a}\omega} V_{\omega,a}^{-1} =
    G_{\omega}\,,
\qquad 
   G_\omega \xi_\omega = \xi_\omega\,.
$$

\noindent The field $\AG$ is represented by a continuous field $\pi= (\pi_\omega)_{\omega\in \Omega}$ of representations, for which 
$\xi$ is a field of cyclic vectors, and satisfying the covariance condition
$$V_{\omega,a}\pi_{\tri^{-a}\omega}(A)V_{\omega,a}^{-1} =
    \pi_{\omega}(\eta_{\omega,a}(A))\,,
\qquad
   A\in\AG(\tra^{-a}\omega)\,.
$$

\noindent The dynamic gives rise to a strongly continuous group of unitary operators on $\Hh_\omega$ with generator $F_\omega$. The 
latter defines a continuous covariant field of self-adjoint operators satisfying 
$$V_{\omega,a} F_{\tri^{-a}\omega} V_{\omega,a}^{-1} =
   F_{\omega}\,,
\qquad 
   F_\omega \xi_\omega = 0\,.
$$

\noindent The family of Tomita-Takesaki modular operators define the covariant continuous field of KMS-dynamics as follows,
$$\theta_t = \alpha_{t\beta}^{(\omega)}\,,\;\;
(t\in\RM)\,,
\quad
  \Delta_\omega
       \prod_{x\in\Lambda} \pi_\omega(s_x a_x^{\sharp_x})
                 \xi_\omega= 
 e^{\beta\sum_{x\in\Lambda} \sharp_x s_x(\epsilon_x -\mu)}
       \prod_{x\in\Lambda} \pi_\omega(s_x a_x^{\sharp_x})
                 \xi_\omega\,,
$$

\noindent where $\sharp$ denotes $\{\cdot, \dag\}$ in the exponent of the $a$'s, while it corresponds to $\{-1,+1\}$ respectively in the exponential. 
The Araki-Connes cones $\Hh_{\omega,+}$ gives also a continuous covariant field of self-dual, homogeneous, oriented cones generated by vectors of the form
$$\pi_\omega\left(
    e^{\beta/4 (H_\Lambda(\omega)-\mu N_\Lambda(\omega))} A 
      e^{-\beta/4 (H_\Lambda(\omega)-\mu N_\Lambda(\omega))}
            \right)\xi_\omega\,,
\quad
  \Lambda\subset \Ll\,, A\in \AG_\Lambda(\omega)\,,\; A\geq 0\,.
$$
\end{proposition}
\noindent Since the proof is straightforward it will be left to the reader.

\begin{rem}
\label{mott09.rem-freeEn}
The notation $F_\omega$ for the generator of the dynamics in $\Hh_\omega$ is justified, because it corresponds to the {\em free energy} 
and can be seen as an infinite volume limit of $H_\Lambda-\mu N_\Lambda$. 
\end{rem}


\section{Dissipative Dynamics}
\label{mott09.sect-dissdyn}

\noindent The general dissipative operator $\DG_\omega= \DG^{kin}_\omega+\DG^\star_\omega$ 
defined in Subsection~\ref{mott09.ssect-model} is the main object of study of this section. 
$\DG$ can be considered as a field of operators $(\DG_\omega)_{\omega\in\Omega}$ on the field of \Css $(\AG(\omega))_{\omega\in\Omega}$. 
Theorem~\ref{mott09.th-close} is the main result of the section where the Friedrich extension Theorem is used to prove that 
the closure of the operator $\DG_\omega$ is a positive self-adjoint operator on the Hilbert space of the GNS representation 
of the algebra of the observables using the equilibrium state.


  \subsection{Complete Positivity}
  \label{mott09.ssect-CP}

\noindent If $\AG$ is a unital \Cs then a map $\eta: \AG \to \AG$ is positive whenever $A\geq 0\Rightarrow \eta(A) \geq 0$. Then $\eta$ extends to $\AG\otimes M_n(\CM)$ by $\eta_n=\eta\otimes id$. Then $\eta$ is called completely positive, (a term which was introduced in \cite{St55}), if $\eta_n$ is positive for all $n$'s. $\eta$ is called
 normalized if $\eta(\id_{\AG})=\id_{\AG}$, whenever $\id_{\AG}$ denotes the unit of $\AG$. By $\CP(\AG)$ 
we will denote the set of completely positive maps and by $\CP_1(\AG)$ the subset of normalized CP-maps.
Examples of CP-maps are:

\begin{enumerate}
\item Any $\ast$-homomorphism is CP. It is normalized if it sends $\id$ to itself. 
\item If $L\in\AG$ then the map $A\mapsto L^\ast AL$ is CP. It is normalized if and only if $L$ is an isometry.
\item If $\Phi$ is CP and $\Phi(\id)$ is invertible, then the map $\Phi'(A) = \Phi(\id)^{-1/2}\Phi(A)  \Phi(\id)^{-1/2}$ is CP and normalized.
\item Any convex combination of CP-maps is CP and the same holds for $\CP_1$. 
\item The composition of two CP-maps is CP and the same holds for $\CP_1$. 
\item If $(\Phi_n)_{n\in\NM}$ is a sequence of CP-maps such that $\Phi(A)= \lim_{n\rightarrow\infty}\Phi_n(A)$ exists for all $A\in\AG$, 
then $\Phi$ is CP, the same holds for $\CP_1$.
\item If $L\in \AG$ let $\Psi_L(A) = L^\ast A+ AL$. Then $e^{-t\Psi_L}(A) = e^{-tL^\ast} A e^{-tL}$ is CP for $t>0$.
\end{enumerate}

\noindent As a result we get the following.

\begin{proposition}
\label{mott09.prop-CPLind}
Let $\AG$ be a $\ZM_2$-graded \Cs with grading automorphism $\sigma$. Let $\Psi:\AG\mapsto\AG$ be a linear map of the form
\begin{equation}
\label{mott09.eq-grLind}
\Psi(A) = 
   \sum_{i=1}^N \Big\{
    \frac{1}{2} \left(L_i^\ast L_i A+ AL_i^\ast L_i\right) - 
     (-1)^{d_{L_i}d_A}L_i^\ast A L_i\Big\}\,,
\end{equation}
\noindent where $L_i\in \AG$ for all $i$'s and all of them are either even or odd elements. 
Then the map $\Phi_t = e^{-t\Psi}$ commutes with the grading and is $\CP_1$ for $t\geq 0$. 
\end{proposition}

\noindent  {\bf Proof: }Since $\Psi$ is a bounded operator, the exponential does exist. Moreover $\Psi(\id)=0$ since the degree of $\id$ is zero, so that $\Phi_t(\id)=\id$ for all $t$'s. Since $\Psi$ is a finite sum of operators, the Trotter product formula \cite{Tro59,Kato} will prove that $\Phi_t\in \CP$ as soon as each pieces of the sum gives a CP-map. Since the grading is given by a $\ast$-automorphism, the map $A\mapsto (-1)^{d_{L_i}d_A}L_i^\ast A L_i$ can be seen as the composition of two CP-maps namely 
$a\mapsto \sigma^d(A)$ (with $d$ the degree of $L_i$) and $A\mapsto L_i^\ast A L_i$. 
Exponentiation is given by a limit of polynomials in these maps with positive coefficients, namely it is CP.
The other maps have the form $A\mapsto L_i^\ast L_i A+ AL_i^\ast L_i$, the exponential of which is CP as well 
(see the last example in the list above).
\hfill $\Box$

\vspace{.1cm}

\noindent The proposition above justifies the form of the Lindbladian in eq.~(\ref{mott09.eq-lind}) for the generator of a Markov semigroup acting on a finite \CS. It also gives the extension to $\ZM_2$-graded \Css.

\vspace{.1cm}

\noindent Let now $\rho$ be a grading invariant state on $\AG$. Then the GNS-construction gives a Hilbert space $\Hh=L^2(\AG,\rho)$, a representation of $\AG$ and a cyclic vector $\xi_0$. Moreover, the state $\rho$ is $\sigma$-invariant, so that $\sigma$ defines a unitary operator $G$ on $\Hh$, called the {\em degree}, such that $G\xi_0 =\xi_0$. Then clearly $G\pi(a) G^{-1} = \pi(\sigma(a))$. In addition, since $\sigma$ is an involution, $G^2=\id_\Hh$, $G^\ast = G$. Let $\Delta,J$ be the modular operator and the modular involution (see Section~\ref{mott09.ssect-gns}) and let $\Hh_+$ be the Araki-Connes homogeneous self-dual oriented cone in $\Hh$.

\vspace{.1cm}

\noindent If $\AG=M_n(\CM)$  and if $\rho$ is the normalized trace $\tr_n$ then the Hilbert space $\Hh$ is the space $L^2(M_n)$ of 
Hilbert-Schmidt operators on $\CM^n$. It can be seen as the set of families $(x_{ij})_{1\leq i,j \leq n}$ with inner product 
$\langle x|y\rangle= (1/n)\sum_{i,j} \overline{x_{ij}} y_{ij}$. The corresponding modular operator is trivial while the conjugacy is 
defined by $j_n(x)_{ij} = \overline{x_{ji}}$. The positive cone is the set $\HS_+(n)$ of positive $n\times n$ Hilbert-Schmidt matrices. 
Consequently, $M_n(\AG)= \AG\otimes M_n(\CM)$ can be endowed with the states $\rho\otimes \tr_n$, leading to the Hilbert space $\Hh\otimes L^2(M_n)$, 
made of families $(x_{ij})_{1\leq i,j \leq n}$ with $x_{ij}\in \Hh$. The inner product is $\langle x|y\rangle_n = (1/n)\sum_{ij} \langle x_{ij}|y_{ij}\rangle$.
 The modular operator is then $\Delta_n= \Delta\otimes \id_n$, with modular conjugacy $J\otimes j_n$, namely $J_n(x)_{ij} = J(x_{ji})$.
 The corresponding positive cone will be denoted by $\Hh_{n+}$.

\begin{definition}
\label{mott09.def-cpH}
A bounded linear map $F$ on $\Hh$ is positivity preserving if $F(\Hh_+)\subset \Hh_+$. It will be called completely positive if $F_n = F\otimes \id_n$ is positivity preserving on $\Hh_n$ for any $n$.
\end{definition}

\noindent It follows from this definition that if $(F_m)_{m\in \NM}$ is a sequence of completely positive maps on $\Hh$ converging weakly to $F$, then $F$ is completely positive as well.

 \subsection{Markov Semi-Groups}
 \label{mott09.ssect-markov}

\noindent One important property of the Lindbladian is the analog of the Leibniz formula for 
the second derivative, namely $(fg)''-f''g-fg''=2 f'g'$. The following formula shows that a Lindbladian 
as defined in eq.~ (\ref{mott09.eq-grLind}) behaves like the analog of $-\Delta$ if 
$\Delta$ is a Laplacian.
For the statement of this proposition, if $\AG$ is a $\ZM_2=\ZM/2\ZM$-graded CAR algebra,
define the  {\em graded commutator} by
\begin{equation}
\label{mott09.eq-grcom}
[A,B]_g = AB- (-1)^{d_A d_B}BA\, \quad \mbox{ for } A, B \in \AG \,.
\end{equation}
\noindent A {\em graded derivation} of degree $d\in \ZM_2$, is a linear operator on the CAR-algebra such that
\begin{equation}
\label{mott09.eq-grder}
\delta(AB)= \delta(A) B + (-1)^{d\cdot d_A}A\; \delta(B)\,.
\end{equation}
\noindent In particular, if $X$ is an element of the CAR-algebra, the map $\delta_X: A\mapsto [X,A]_g$ is a {\em graded derivation} with degree $d_X$.

\begin{proposition}
\label{mott09.prop-leibniz}
Let $\LG$ be the Lindblad operator given by $\LG(A)=\imath [F,A]+\Psi$ where $F=F^\ast$ and
$\Psi$ is given as in eq.~(\ref{mott09.eq-grLind}). Then for any pair $A,B\in\AG$

\vspace{.2cm}

\noindent (i) $\LG(A^\ast)= \left(\LG(A)\right)^\ast$

\vspace{.2cm}

\noindent (ii) $\LG(A^\ast B) -A^\ast\LG(B) -\LG(A^\ast)B = -\sum_{i=1}^N [L_i,A]_g^\ast [L_i,B]_g$~~~~~~~~ {\bf(Leibniz formula)}.
\end{proposition}

\vspace{.2cm}

\noindent {\bf Proof: }(i) Since $F=F^\ast$, the first claim is obvious by inspection. 

\vspace{.1cm}

\noindent (ii) The map $\delta: A\in\AG \mapsto \imath [F,A]$ is a $\ast$-derivation and therefore it 
satisfies the Leibniz formula, 
$\delta(A^\ast B)= \delta(A)^\ast B + A^\ast \delta(B)$. Hence this part does not contribute 
to the {\em r.h.s.}. It is enough then to consider 
the case $N=1$. Let J denote the left hand side of the Leibniz formula for $N=1$. Then
\begin{eqnarray*}
2J & =&L^\ast L A^\ast B+A^\ast B L^\ast L 
       -A^\ast L^\ast L B - A^\ast B L^\ast L 
        - L^\ast L A^\ast B - A^\ast L^\ast L B\\
&& -2(-1)^{d_L (d_A+d_B)} L^\ast A^\ast B L
    +2(-1)^{d_L d_B} A^\ast L^\ast  B L+
     2(-1)^{d_L d_A} L^\ast A^\ast L  B\,.
\end{eqnarray*}
\noindent After cancellation this gives
$$2J= -2A^\ast L^\ast \Big(L B -(-1)^{d_L d_B}B L\Big)
    +2(-1)^{d_L d_A} L^\ast A^\ast \Big(LB - (-1)^{d_L d_B} B L\Big)\,,
$$
\noindent leading to
$$2J= -2\Big(LA - (-1)^{d_L d_A}AL \Big)^\ast
       \Big(L B -(-1)^{d_L d_B}B L\Big)
    = -2[L,A]_g^\ast [L,B]_g\,.
$$
\hfill $\Box$

 \subsection{A Convergence Theorem}
 \label{mott09.ssect-conv}

\noindent In this section a convergence result will be obtained for Lindblad operators with an infinite number of jump operators acting on the 
algebra $\AG(\omega)$. It is worth noticing that the proof of this result is modeled on a similar result found in \cite{BrRo2}. 
For each finite set $X\subset \Ll$  let $L_X$ be a continuous section of the field $\AG$ such that
\begin{enumerate}
  \item $L_X(\omega)\in \AG_X(\omega)$ and has degree $d_X$,
  \item the section is covariant, namely $\eta_{\omega,a}\left(L_{X-a}(\tra^{-a}\omega)\right) = L_X(\omega)$,
  \item for all natural numbers $N\in\NM$ there exists  $p>0$ such that 
\begin{equation*}
\label{mott09.eq-normDG}
\sum_{m\in\NM} e^{pm}\; \sum_{n=0}^N\;
   \sup_{\omega\in\Omega}\;
    \sum_{0\in X\,;\,\diami(X)=m;|X|=n+1}
     \|L_X(\omega)\|^2 =: C_L <\infty\,,
\end{equation*}
\noindent where $\diam(X)$ denotes the diameter of the set $X$ and $|X|$ the number of points in $X$. For convenience, the diameter will be 
calculate w.r.t. the metric in $\RM^d$ given by $|x|=\max{|x_i|\,;\, 1\leq i\leq d}$. Hence, a ball of diameter $R$ is a cube of side $R$ with sides 
parallel to the canonical basis of $\RM^d$. Such a cube has a volume $R^d$. On the other hand, only the nonempty $X$'s matter, so that $|X|=n+1$ 
for some natural integer $n$. 
\end{enumerate}

\noindent Then $\Dd_{X,\omega}$ denotes the operator acting on $\AG(\omega)$ defined by
$$
\Dd_{X,\omega}(A) =
   \frac{1}{2}\{L_X(\omega)^\ast L_X(\omega),A\} - 
    (-1)^{d_Xd_A}L_X(\omega)^\ast A L_X(\omega)\,,
$$
\noindent where $\{A,B\}= AB+BA$ is the anti-commutator. Then the following result holds.

\begin{theorem}
\label{mott09.th-convAG}
Let $\fs$ denote the set of finite subsets of $\Ll$. Then, the operator $\DG_\omega= \sum_{X\in\fs}\;\Dd_{X,\omega}$ is well defined on $\AGl(\omega)$. 
It is covariant and, if $\xi$ is a continuous section of $\AGl$, its image $\DG(\xi)$ is a continuous section of $\AG$. In addition its exponential 
$e^{-t\DG_\omega}$ defines a continuous covariant field of Markov semi-groups.
\end{theorem}

\noindent {\bf Proof: } 1)- Let $A\in\AG_\Lambda(\omega)$. Then $\Dd_{X,\omega}(A)=0$ whenever $X\cap\Lambda=\emptyset$. Hence the sum over $X$ is restricted 
to those subsets $X$ intersecting $\Lambda$. From the estimate given in eq.~(\ref{mott09.eq-normDG}), it follows easily that the sum defining $\DG_\omega(A)$ 
converges in norm, uniformly {\em w.r.t.} $\omega$. Then the continuity and the covariance are straightforward to check.

\vspace{.1cm}

\noindent 2)- Iterating the definition of $\DG$ gives (omitting $\omega$)
$$\DG^k(A) =
   \sum_{X_1,\cdots, X_k}
     \Dd_{X_k}\circ\cdots\circ \Dd_{X_1}(A)\,,
$$
\noindent where the family $(X_1,\cdots,X_k)$ of finite subsets of $\Ll$ is submitted to satisfy the {\em compatibility condition} 
$X_j\cap\Lambda_{j-1}\neq \emptyset$ for $1\leq j\leq k$, whenever $\Lambda_0=\Lambda$ and $\Lambda_j= \Lambda_{j-1}\cup X_j$. It is easy to check that
$$\|\Dd_{X}(A)\| \leq
   2\|L_X\|^2 \|A\|\,.
$$
\noindent Therefore, the {\em l.h.s.} can be estimated by
$$\|\DG^k(A)\| \leq \|A\|
   \sum_{X_1,\cdots, X_k} 
    2^k \prod_{j=1}^k \|L_{X_j}\|^2\,.
$$
\noindent By assumption, $\|L_{X_j}\|^2\leq C_L \,e^{-p\,\diami(X_j)}$. Let $N(m_1,n_1;\cdots;m_k,n_k)$ denote the number of compatible families 
$(X_1,\cdots,X_k)$ such that $\diam(X_j)=m_j$ and $|X_j|=n_j+1$. This gives
$$\|\DG^k(A)\| \leq (2C_L)^k 
   \sum_{m_1,\cdots, m_k} e^{-p(m_1+\cdots+ m_k)} 
    \sum_{n_1,\cdots, n_k}
     N(m_1,n_1;\cdots;m_k,n_k)\,.
$$
\noindent In order to estimate $N(m_1,n_1;\cdots;m_k,n_k)$, it ought to be remarked that 

\vspace{.2cm}

\noindent (i) $|X|\leq \diam(X)^d$, so that $n_j< m_j^d$ for all $j$,

\noindent (ii) the maximal number of choices for $X_j$ is obtained by choosing a point in $\Lambda_{j-1}$ then by choosing $n_j$ points in a
 hypercube of side at most $m_j$. There is at most $m_j^{dn_j}\times |\Lambda_{j-1}|$ ways of making this choice.

\noindent (iii) the number of points in $\Lambda_{j-1}$ is at most $\{|\Lambda|+ n_1+ n_2+\cdots n_{j-1}\}\leq \{|\Lambda|+N(j-1)\} \leq Nj(1+ (|\Lambda|/N-1)/j)$,
 by construction. In particular this gives
$$N(m_1,n_1;\cdots;m_k,n_k) \leq N^k\;k!\;
   \prod_{j=1}^k m_j^{dN}\;
    \prod_{j=1}^k \left(1+ \frac{|\Lambda|-N}{Nj}\right)\,.
$$
\noindent Then using $(1+u)\leq e^u$ and also $1+1/2+\cdots + 1/k \leq 1+\ln(k)$ gives
$$N(m_1,n_1;\cdots;m_k,n_k) \leq N^k\;k!\;
  e^{|\Lambda|/N-1}\; k^{|\Lambda|/N-1}\;
   \prod_{j=1}^k m_j^{dN}\;
$$
\noindent It follows that
$$\sum_{k=0}^\infty |t|^k\;
   \frac{\|\DG^k(A)\|}{k!} \leq \|A\|\;
    \sum_{k=0}^\infty k^{\kappa}(C_1|t|)^k
$$
\noindent with $\kappa =|\Lambda|/N-1$ and $C_1 =2NC_L \sum_{m\geq 1} m^{dN}e^{-pm}<\infty$. Hence, if $C_1|t|<1$, the sum defining $e^{-t\DG}(A)$
 converges absolutely and uniformly in $\omega\in\Omega$. Since $C_1$ does not depend on the volume $\Lambda$ it follows that $e^{-t\DG}$ is well 
defined on $\AG$ for $t\in\CM$ such that $|t|<C_1^{-1}$.

\vspace{.1cm}

\noindent 3)- Using the Proposition~\ref{mott09.prop-CPLind}, and the previous convergence, it follows that $e^{-t\DG}$ is $\CP$ for $0\leq t <C_1^{-1}$. 
Moreover, since $\DG(\id)=0$, it is actually $\CP_1$. In particular, it is a contraction semi-group. Moreover, by construction of the exponential, 
if $s,t\in\CM$ are such that $|s|+|t|< C_1^{-1}$, then  $e^{-(s+t)\DG} = e^{-s\DG}\,e^{-t\DG}$. Therefore if $t\in\RM_+$ let $n$ be an integer such 
that $t/n < C_1^{-1}$. Then  $e^{-t\DG}= \left(e^{-t/n\DG}\right)^n$ is well defined and does not depend upon which $n$ has been chosen. And for 
the same reason it defines a Markov semi-group. 

\vspace{.1cm}

\noindent 4)- The continuity and the covariance follow from the definition and the proof will be left to the reader.
\hfill $\Box$


 \subsection{Jump Dynamics}
 \label{mott09.ssect-jumpdy}

\noindent The general dissipation operator $\DG_\omega$ defined in Section~\ref{mott09.ssect-model} is the main object of study in this subsection. The main result is Theorem~\ref{mott09.th-close}.

\vspace{.1cm}

\noindent The model defined in Section~\ref{mott09.ssect-model} is a specific example of a larger class of models of the form
\begin{equation}
\label{mott09.eq-dissiJump}
\DG_\omega(A) =
   \sum_{\gau \in \js}\left(
    \frac{1}{2}\{ L_\gau^\ast(\omega) L_\gau(\omega),A\}
     -(-1)^{d_{L_\gau}d_A}
      L_\gau^\ast(\omega) A L_\gau(\omega)
     \right)\,,
\end{equation}
\noindent where the following axioms are satisfied.
\begin{itemize}
\item[J1-]The index set $\js$, called the set of {\em jumps}, is countable and admits a bijective involution $\gau\in\js\mapsto \gao\in\js$ called 
{\em time-reversal}. Moreover, the translation group $\ZM^d$ acts on $\js$ in a bijective way and the action is denoted by $\gau\mapsto \gau+a$ by 
mappings commuting with the involution.

\item[J2-] For each $\gau\in\js$ there is a continuous covariant field of local observables $L_\gau$, called the {\em jump operators}. 
In particular\\ (i) $\exists \,\Lambda \subset \Ll$ finite, depending on $\gau$, so that $L_\gau(\omega)\in \AG_\Lambda(\omega)\,,\;\forall\omega\in\Omega$,\\
(ii)  the smallest such $\Lambda$ is called the {\em support} of $\gau$ and is denoted $\supp\{\gau\}$ and satisfies $\supp\{\gau+a\} = \supp\{\gau\} +a$,\\
(iii) $\eta_{\omega,a}\left\{L_{\gamma-a}(\tra^{-a}\omega)\right\} = L_{\gau}(\omega)$ and $L_\gau\,,\,L_{\gau -a}$ have the same degree.

\item[J3-] Under the time evolution the jump operators satisfy $\alpha_t(L_\gau) = e^{\imath t\varepsilon_\gau} \; L_\gau$ for all $t\in\RM$. 

\noindent where the $\varepsilon_\gau :\Omega\mapsto \RM$'s are continuous functions.

\item[J4-] The jump operators satisfy a {\em $\beta$-KMS condition}, namely $L_\gau^\ast = e^{-\beta \varepsilon_{\gau}/2}\;L_{\gao}$, in particular 
$\varepsilon_{\gao}=-\varepsilon_\gau$ and $L_{\gao}$ has the same degree as $L_\gau$. 

\item[J5-] The following sum converges $\sum_{\gamma\,;\,0\in \supi\{\gamma\}} L_\gau^\ast(\omega) L_\gau(\omega) \in \AG(\omega)$ uniformly
 with respect to $\omega$. 

\end{itemize}

\begin{rem}
\label{mott09.rem-bKMS}
The $\beta$-KMS condition is also called {\em detailed balance} in the Physics literature (see for instance \cite{Spohn80}). This axiom is essential to describe the dissipative evolution {\em at equilibrium}.
\end{rem}
\begin{proposition}
\label{mott09.prop-disExists}
If the jump operators satisfy the assumptions [J1-J5] {\rm (}except possibly  the axiom [J4]{\rm )}, the operator $\DG_\omega$ is well defined by eq.~(\ref{mott09.eq-dissiJump}) on the set of local observables. In addition
\begin{align}
\alpha_t^{(\omega)}\circ \DG_\omega \circ \alpha_{-t}^{(\omega)} =
   \DG_{\omega}\,,
\hspace{3cm}
&\mbox{\rm\bf (time-invariance)}  \\
\eta_{\omega,a}\circ \DG_{\tri^{-a}\omega} \circ \eta_{\omega,a}^{-1} =
   \DG_{\omega}\,,
\hspace{3cm}
&\mbox{\rm\bf (space-covariance)}
\end{align}
\noindent and, if $\xi\in\fs$, the field $\omega\in\Omega\mapsto \DG_\omega(\xi(\omega))$ is continuous. 
\end{proposition}

\noindent {\bf Proof: } Let $A$ be localized in $\Lambda_0$. Each $L_\gau$ belongs to the local algebra $\AG_{\supi\{\gamma\}}$.
 In particular if $\supp\{\gamma\}\cap \Lambda_0 =\emptyset$, it follows that $A$ commutes with $L_\gau^\ast L_\gau$ and that 
$(-1)^{d_{L_\gau}d_A} L_\gau^\ast A L_\gau = L_\gau^\ast L_\gau A$. Thus, the term 
$\frac{1}{2}\{ L_\gau^\ast L_\gau,A\}-(-1)^{d_{L_\gau}d_A} L_\gau^\ast A L_\gau$ just vanishes. Hence,
$$\DG_\omega (A) =
   \sum_{\gau \in \js\,;\,\supi\{\gamma\}\cap\Lambda_0\neq \emptyset}
    \left(
     \frac{1}{2}\{ L_\gau^\ast L_\gau,A\}-(-1)^{d_{L_\gau}d_A}
      L_\gau^\ast A L_\gau
   \right)\,.
$$
\noindent Thanks to the axiom [J5], this sum converges in $\AG$. Thanks to axiom [J2-J3] the time-invariance, the 
covariance condition and the continuity are satisfied.
\hfill $\Box$

\begin{proposition}
\label{mott09.prop-DGsad}
Let $\rho$ be the continuous field of $\beta$-KMS states over $\AG$ defined in Prop.~\ref{mott09.prop-gs}. If the jump operators satisfy [J1-J5],
 then $\rho$ is $\DG$-invariant and satisfies for all finite $\Lambda \subset \Ll$ and all $A,B\in\AG_\Lambda(\omega)\,$
\begin{equation}
\rho_\omega\left(A^\ast \DG_\omega(B)\right) =
   \rho_\omega\left(\DG_\omega(A)^\ast B\right)=
     \frac{1}{2}
    \sum_{\gau\in\js}
     \rho_\omega\left(
        [L_\gau(\omega),A]_g^\ast [L_\gau(\omega),B]_g 
                \right)\;.
\end{equation}
\end{proposition}

\begin{rem}
\label{mott09.rem-Lap}
This shows that $\DG$ acts on the GNS-representation of the ground state as a a positive self-adjoint operator. Moreover it 
can be seen as a generalization of a Laplacian. For indeed, the family of (graded) commutators by $L_\gau$, indexed by $\gau\in\js$, can be 
seen as a gradient and the right hand side of the previous equation looks like a Sobolev norm of the type $\int |\nabla A|^2$ if $A=B$. 
\end{rem}

\noindent {\bf Proof: } Since there is no confusion, the reference to $\omega\in\Omega$ will be omitted. Thanks to the Leibniz formula 
(Proposition~\ref{mott09.prop-leibniz}), it is sufficient to show that (i) $\rho\circ\DG=0$ ($\DG$-invariance of $\rho$) and (ii) $\DG$ is symmetric. 

\vspace{.1cm}

\noindent 1)- Let $A\in\AG_\Lambda$. Since $\DG(A)$ has the same degree as $A$, it is sufficient to assume that $d_A=0$, because $\rho(B)=0$ 
whenever $B$ has degree one. Then $\rho\left(\DG(A)\right)$ is the sum of three types of terms, namely 
$(1/2)\rho\left(L_\gau^\ast L_\gau A\right)$, $(1/2)\rho\left(AL_\gau^\ast L_\gau \right)$  and $(-1)\rho\left(L_\gau^\ast A L_\gau\right)$. 
Thanks to [J3], it follows that $L_\gau^\ast L_\gau$ is invariant by the dynamics $\alpha_t$. In particular 
$\alpha_{-\imath\beta}(L_\gau^\ast L_\gau) = L_\gau^\ast L_\gau$. Thus, tanks to the $\beta$-KMS condition (eq.~(\ref{mott09.eq-KMS}))
$$\rho(A\,L_\gau^\ast L_\gau) =
   \rho\left(\alpha_{-\imath \beta}(L_\gau^\ast L_\gau) A\right) =
    \rho\left(L_\gau^\ast L_\gau A\right)\,.
$$
\noindent In particular the first two terms are equal. Moreover, the axioms [J3-J4] imply that 
$\alpha_{-\imath\beta}(L_\gau)= e^{\beta\varepsilon_\gau} L_\gau= e^{\beta\varepsilon_\gau/2} L_{\gao}^\ast$ and, similarly, 
$e^{\beta\varepsilon_\gau/2} L_\gau^\ast = L_{\gao}$. Hence
$$\rho(L_\gau^\ast AL_\gau) =
   \rho(\alpha_{-\imath\beta}(L_\gau)L_\gau^\ast A) =
    \rho(L_{\gao}^\ast L_{\gao}\,A) \,. 
$$
\noindent Therefore, since the time-reversal $\gau\mapsto \gao$ is a bijection, the sum of these last terms compensate the sum of the 
other terms to give zero.

\vspace{.1cm}

\noindent 2)- To prove that $\DG$ defines a symmetric operator, let $A,B$ be elements in $\AG_\Lambda$. Without loss of generality, 
it can be assumed that $A$ and $B$ have the same degree, otherwise the $\rho$-average vanishes and the identity becomes trivial. Then 
$\rho(A^\ast \DG(B))$ is a sum of three types of terms. The first ones are 
$$\rho(A^\ast L_\gau^\ast L_\gau B) = 
   \rho\left( (L_\gau^\ast L_\gau A )^\ast B\right)\,.
$$
\noindent The next terms have the form 
$$\rho(A^\ast B L_\gau^\ast L_\gau ) = 
   \rho\left(L_\gau^\ast L_\gau A^\ast B\right) = 
    \rho\left((A L_\gau^\ast L_\gau)^\ast B\right)\,,
$$
\noindent where the middle identity comes from the $\beta$-KMS condition and the invariance of $L_\gau^\ast L_\gau$ under the time evolution 
(axiom  [J3]). The last terms are more involved. Using again the $\beta$-KMS condition, the identities $d_A=d_B$, $d_\gau=d_{\gao}$ and the axiom [J4], leads to
$$(-1)^{d_\gau d_B}\rho(A^\ast L_\gau^\ast B L_\gau) = 
   (-1)^{d_\gau d_A}  e^{\beta\varepsilon_\gau} 
    \rho\left(L_\gau A^\ast L_\gau^\ast B\right) = 
     (-1)^{d_{\gao} d_A}
      \rho\left((L_{\gao}^\ast A L_{\gao})^\ast B\right)\,,
$$
\noindent Since the map $\gau \mapsto \gao$ is a bijection the sum of all these terms gives $\rho(A^\ast \DG(B))= \rho(\DG(A)^\ast B)$.
\hfill $\Box$

\vspace{.2cm}

\noindent Through the GNS-representation, the field of ground-states $\rho$ defines a continuous translation covariant field of Hilbert spaces 
$\Hh_\omega = L^2(\AG(\omega), \rho_\omega)$ \cite{Kad,Ta03a}. $\Hh_\omega$ is obtained from $\AG(\omega)$ through the inner product 
$\langle A|B\rangle_{\omega}=\rho_\omega(A^\ast B)$, after taking the quotient by the subspace of elements of zero norms and completing. The 
canonical map from $\AG(\omega)$ into $\Hh_\omega$ will be denoted by $\zeta_\omega$. Hence 
$$\rho_\omega(A^\ast B)=
   \langle \zeta_\omega(A)| \zeta_\omega(B)\rangle_{\omega}\,,
\quad
    \rho_\omega(A^\ast A)=\|\zeta_\omega(A)\|_{\omega}^2\,,
\qquad
    A,B\in \AG(\omega)\,.
$$
\noindent On each element of this field the Proposition~\ref{mott09.prop-DGsad} defines a densely defined field of positive quadratic forms as follows
\begin{equation}
\label{mott09.eq-qform}
\Qq_\omega (A,B) = 
     \frac{1}{2}
    \sum_{\gau\in\js}
     \rho_\omega\big(
        [L_\gau(\omega),A]_g^\ast [L_\gau(\omega),B]_g 
                \big)\,.
\end{equation}
\noindent The following result shows that, as a consequence of the Friedrich extension theorem \cite{RS}, this form defines a positive self adjoint 
operator $\Hh_\omega$.

\begin{theorem}
\label{mott09.th-close}
If the jump operators satisfy [J1-J5], the quadratic form $\Qq_\omega$, which is densely defined on $\Hh_\omega$, is closable. Its closure defines a
 positive self-adjoint operator, denoted by $\ds_\omega$, on $\Hh_\omega$. The contraction semi-group $e^{-t\ds_\omega}$ is completely positive. The 
corresponding field of contraction semi-groups is continuous, time-invariant and covariant.
\end{theorem}

\noindent {\bf Proof: } (i) {\em Closability.} By abuse of notation $A$ will represent here either an element of $\AG(\omega)$ or its image in 
$\Hh_\omega$. Let $\|A\|_{\omega,Q}$ denote the norm 
$$\|A\|_{\omega,Q}^2 =
   \|A\|_\omega^2 + \Qq_\omega (A,A)
$$
\noindent To prove that $\Qq_\omega$ is closable, let $(A_n)_{n\in\NM}$ be a $\|\cdot \|_{\omega,Q}$-Cauchy sequence in $\AGl(\omega)$ such that 
$\lim_{n\rightarrow \infty}\|A_n\|_\omega =0$. It should be proved that $\lim_{n\rightarrow \infty}\|A_n\|_{\omega,Q} =0$. Since this sequence is 
$\|\cdot \|_{\omega,Q}$-Cauchy, it follows that $\Qq_\omega(A_n,A_n)$ converges and is therefore uniformly bounded in $n$. In particular, it follows 
that $\delta(A_n) = \left([L_\gau(\omega),A_n]_g \right)_{\gau\in\js}$ is Cauchy if seen as an element of the Hilbert space 
$\Hh_\omega\otimes \ell^2(\js)$. Therefore there is $\xi= (\xi_\gau)_{\gau\in\js}\in \Hh_\omega\otimes \ell^2(\js)$, such that $\delta(A_n)\rightarrow \xi$. 
Hence, given $\epsilon >0$ there is a finite subset $I\subset \js$, such that $\sum_{\gau\notin I} \|\xi_\gau\|_{\omega}^2 \leq \epsilon$. Now, thanks 
to [J1-J5], and omitting $\omega$ inside $L_\gau$
\begin{eqnarray*}
\rho_\omega \left(|[L_\gau, A_n]_g|^2\right) & = &
 \rho_\omega \left(A_n^\ast L_\gau^\ast  L_\gau A_n\right) + 
 \rho_\omega \left(L_\gau^\ast A_n^\ast A_n L_\gau\right)\\
& & - (-1)^{d_{L_\gau}d_{A_n}}
\Big(
\rho_\omega \left(A_n^\ast L_\gau^\ast  A_n L_\gau\right) +
\rho_\omega \left(L_\gau^\ast A_n^\ast  L_\gau A_n\right)
\Big)\,.
\end{eqnarray*}

\noindent The last two terms of the {\em r.h.s.} can be estimated in terms of the first two. Since $ L_\gau(\omega)$ is bounded, the first term is 
bounded by $\| L_\gau(\omega)\|^2 \|A_n\|_\omega^2$ which converges to zero. Using the axiom [J3-J4], the second term can be written (omitting $\omega$), as
$$\rho \left(L_\gau^\ast A_n^\ast A_n L_\gau\right)=
   e^{\beta\varepsilon_\gau}
    \rho \left(L_\gau L_\gau^\ast A_n^\ast A_n \right)= 
    \rho \left(L_{\gao}^\ast L_{\gao} A_n^\ast A_n \right) \,.
$$
\noindent Thanks to [J3], it follows that $L_{\gao}^\ast L_{\gao}$ is invariant by the modular automorphism, so that, using the Cauchy-Schwartz inequality, 
$$\rho \left(L_{\gao}^\ast L_{\gao} A_n^\ast A_n \right) \leq 
   \|A_n\| 
    \rho \left(
     L_{\gao}^\ast L_{\gao} A_n^\ast A_n L_{\gao}^\ast L_{\gao}
         \right)^{1/2}=
  \|A_n\| 
    \rho \left(
     \big(L_{\gao}^\ast L_{\gao}\big)^2 A_n^\ast A_n
         \right)^{1/2}\,.
$$
\noindent Iterating $m$-times, leads to 
$$\rho \left(L_{\gao}^\ast L_{\gao} A_n^\ast A_n \right) \leq
  \|A_n\|^{2-2^{-m}}
    \rho \left(
     \big(L_{\gao}^\ast L_{\gao}\big)^{2^{m-1}} A_n^\ast A_n
         \right)^{2^{1-m}}\leq
   \|A_n\|^{2-2^{-m}} \|L_{\gao}\|^2 \|A_n\|_{\AG}^{2^{-m}}\,.
$$
\noindent Consequently, as $m\rightarrow \infty$,
$$\rho \left(L_{\gao}^\ast L_{\gao} A_n^\ast A_n \right) \leq
  \|A_n\|^{2}\|L_{\gao}\|^2 \,,
$$
\noindent which is also converging to zero as $n\rightarrow \infty$. Hence, the finite sum $\sum_{\gau\in I} \rho \left(|[L_\gau,A]|^2\right)$ 
vanishes, so that $\|\xi\|\leq \epsilon$. Since $\epsilon$ can be taken arbitrarily small, it follows that $\xi=0$, proving that $\Qq_\omega$ is closable. 

\vspace{.1cm}

\noindent It follows from the Friedrich extension method that there exists a positive self-adjoint operator $\ds_\omega$, with domain 
$\Dd = \{ A\in \Hh_\omega\,;\, \exists C>0, \| \Qq_\omega \|_\omega (B,A)\leq C \|B\|_\omega \,,\, \forall B\in \Hh_\omega\}$, defined by 
\begin{equation}
\Qq_\omega (B,A) = 
   \langle B|\ds_\omega\,A\rangle_\omega\,,
\qquad
    A\in \Dd\,.
\end{equation}
\noindent (ii) {\em Complete Positivity.} Let $\js$ be finite to begin with. Thanks to the Propositions~\ref{mott09.prop-DGsad} and 
\ref{mott09.prop-CPLind}, it follows that the complete positivity holds if $\js$ is finite in the algebra $\AG(\omega)$. Moreover, since $\DG$ 
commutes with the dynamics (Proposition~\ref{mott09.prop-disExists}), the operator $\ds_\omega$ commutes with the Modular operator $\Delta_\omega$. 
In particular $e^{-t\ds_\omega}$ will be completely positive on $\Hh_\omega$ (see Section~\ref{mott09.ssect-CP}). If $\js$ is not finite, then 
$\Qq_\omega$ can be seen as the supremum of a countable family of similar forms with $\js$ finite. Correspondingly there is a non decreasing 
sequence of positive self-adjoint operators $\ds_{n,\omega}$ converging weakly to $\ds_\omega$. The Lemma~\ref{mott09.lem-QnQ} belows shows 
then that $\ds_{n,\omega}$ converges to $\ds_\omega$ in the strong resolvent sense. This, in turns proves that the semi-group $e^{-t\ds_{n,\omega}}$
 converges strongly to $e^{-t\ds_{\omega}}$. In particular, $e^{-t\ds_{\omega}}$ is completely positive. 

\vspace{.1cm}

\noindent (iii) {\em Covariance and Continuity.} The same argument as before shows that the continuity of the field $\omega\in\Omega \mapsto \Qq_{\omega}$ 
implies the strong resolvent continuity of the field $\omega\in\Omega \mapsto \ds_{\omega}$. Hence the semi-group $e^{-t\ds_{\omega}}$ is also continuous 
in $\omega$. The covariance is a simple consequence of Proposition~\ref{mott09.prop-disExists}.
\hfill $\Box$

\begin{lemma}
\label{mott09.lem-QnQ}
Let $A$ be a positive self-adjoint operator on the Hilbert space $\Hh$, with dense domain $\Dd$. Let $(A_n)_{n\in\NM}$ be a non decreasing sequence 
of bounded positive operators converging weakly to $A$ on $\Dd$. Then the resolvent of $A_n$ converges strongly to the resolvent of $A$.
\end{lemma}

\noindent  {\bf Proof: } Since $0\leq A_1\leq A_2\leq \cdots \leq A_n\leq \cdots\leq A$, it follows that the sequence of bounded operators $(1+A_n)^{-1}$
converges weakly. Let $R$ be the weak limit. Since the sequence is bounded the sequence converges strongly as 
well. Then the inequality 
$1\geq (1+A_n)^{-1}\geq (1+A)^{-1}$ implies $1\geq R \geq (1+A)^{-1}$. In particular,
$$
1+A \geq (1+A)^{1/2}(1+A_n)^{-1}(1+A)^{1/2} \geq (1+A)^{1/2}R(1+A)^{1/2}\geq 1\;.
$$
Hence, $(1+A)^{1/2}R(1+A)^{1/2}$ is invertible and its inverse satisfies 
$$(1+A)^{-1/2}(1+A_n)(1+A)^{-1/2}\leq (1+A)^{-1/2}R^{-1}(1+A)^{-1/2}\leq 1\,.$$ 
Since the left hand side of this inequality converges weakly to $1$, thus also 
strongly, it follows that $R^{-1}= (1+A)$. Thus $s-\lim_{n\rightarrow \infty} (1+A_n)^{-1} = (1+A)^{-1}$. By standard arguments, this shows that for 
$z\in\CM\setminus \RM_+$, \mbox{$s-\lim_{n\rightarrow \infty}(A_n-z)^{-1} = (A-z)^{-1}$}.
\hfill $\Box$


\section{The return to equilibrium}
\label{mott09.sect-model}

\noindent In this section we first justify the introduction of the thermal bath part of the dissipation operator in eq.~(\ref{mott09.eq-dgstar}).
Then it will be shown that the dissipation operator $\LG_\omega$, given in eq.~(\ref{mott09.eq-LindMod}), defines a semigroup on the state space. 
Furthermore we obtain that the point wise limit of the semigroup at any initial state is equal to the equilibrium state as the time parameter tends to infinity (Theorem~\ref{mott09.prop-unInvSt}). 
This justifies the title of the present Section. 
Finally, the spectrum of the dissipation operator $\DG_\omega$ is examined which allows us to conclude that the return to equilibrium as exponentially fast in time.

\vspace{.1cm}

\noindent If $\DG$ is the generator of a semigroup $(e^{-t\DG})_{t \geq 0}$ on a \Cs $\AG$, then an element $A \in \AG$ is called invariant for the semigroup if $A$ is an eigenvector of $e^{-t\DG}$ corresponding to the eigenvalue $1$ for every $t \geq 0$. Also, if $\rho$ is any state on the \Cs $\AG$ then $\rho$ is called $\DG$-invariant if 
$$\rho (e^{-t\DG}(B))= 
   \rho (B) 
\qquad \forall B \in \AG\,,\;\forall t \geq 0 \, .
$$
\noindent The following result states elementary properties of the jump operators which are defined by eq.~(\ref{mott09.eq-Lxy}). It can be obtained by inspection and the proof will be left to the reader.

\begin{proposition}
\label{mott09.prop-x2y}
The jump operators $L_{x\rightarrow y}(\omega)$ given by eq.~(\ref{mott09.eq-Lxy}) \& (\ref{mott09.eq-Gammxy}) have degree zero and satisfy the axiom [J1-J5] with

(i) $\supp\{x\rightarrow y\} = \{x,y\}$,

(ii) $\{x\rightarrow y\} +a= \{x+a\rightarrow y+a\}$

(iii) the time-reversal corresponds to $\{y\rightarrow x\}$,

(iv) $\varepsilon_{x\rightarrow y} = \epsilon_y-\epsilon_x$. 
\end{proposition}

\noindent In order to justify the need for the thermal bath part of the dissipation operator, it is enough to observe that the kinetic part $\DG_\omega^{kin}$ suffers from an annoying disease: since it leaves the number operator invariant, it does not have a unique invariant state.
In fact, varying the chemical potential $\mu$ in \eqref{mott09.eq-rhonx} produces infinitely many invariant states.
Thus, the thermal bath fixes the chemical potential and the
kinetic part of the dissipation operator alone cannot drive the electron gas towards equilibrium. 
This is the content of the following result.

\begin{proposition}
\label{mott09.prop-noUniqEq}
The kinetic part of the dissipation operator, given in eq.~(\ref{mott09.eq-dgkin}), commutes with the number operator. In particular its set of invariant state is not reduced to a point.
\end{proposition}

\noindent {\bf Proof: } The number operator has been defined in eq.~(\ref{mott09.eq-nbodyNH}) for finite volume. As for the Hamiltonian dynamics it 
generates an automorphism group which is defined, in the infinite volume limit by
$$\nu_t^{(\omega)}(s_xa_x) = 
   e^{-\imath t}s_xa_x\,,
\qquad
    \nu_t^{(\omega)}(s_xa_x^\dag) = 
     e^{\imath t}s_xa_x^\dag\,.
$$
\noindent In particular, $L_{x\rightarrow y}(\omega)$ is invariant by this automorphism group. Therefore, like in Proposition~\ref{mott09.prop-disExists}, 
it follows that
$$\nu_t^{(\omega)}\circ \DG_\omega^{kin} \circ \nu_{-t}^{(\omega)} = 
   \DG_\omega^{kin}\,.
$$
\noindent Hence any state generated by a finite volume Hamiltonian of the form 
$$F_\Lambda(\omega) = 
   P\left( \sum_{x\in\Lambda} \epsilon_x n_x\right) + 
    Q\left( \sum_{x\in\Lambda} n_x\right)\,,
$$
\noindent where $P$ and $Q$ are polynomials, is invariant by $\DG_\omega^{kin} $. 
\hfill $\Box$

\vspace{.1cm}

\noindent The following result states elementary properties of the jump operators which are defined by eq.~(\ref{mott09.eq-Lxstar}). It can be obtained by inspection and the proof will be left to the reader.

\begin{proposition}
\label{mott09.prop-x2cem}
The jump operators $L_{x\rightarrow \star}(\omega)$ and $L_{\star\rightarrow x}(\omega)$, given by 
eq. (\ref{mott09.eq-Lxstar}) \& (\ref{mott09.eq-Gammxstar}), have degree one and satisfy the axiom [J1-J5] with

(i) $\supp\{x\rightarrow \star\}= \{x\} = \supp\{\star\rightarrow x\}$

(ii) $\{x\rightarrow \star\}+a=\{x+a\rightarrow \star\}$ and similarly for $\{\star\rightarrow x\}$

(iii) ${x\rightarrow \star}$ is time reversed from ${\star\rightarrow x}$, 

(iv) $\varepsilon_{\star\rightarrow x}= \epsilon_x-\mu$. 
\end{proposition}

\noindent The following theorem shows the return to equilibrium $\rho_\omega$ as defined
in Proposition~\ref{mott09.prop-gs} for the dynamical system 
$(e^{-t\DG_\omega})_{t \geq 0}$.

\begin{theorem}
\label{mott09.prop-unInvSt}
The operators $\DG_\omega^\star$ and $\DG_\omega$ have a unique invariant state given by $\rho_\omega$.
Also, for every state $\tilde{\rho}$ on the \Cs $\AG (\omega )$, and for every observable 
$A \in \AG (\omega )$,
\begin{equation}
\lim_{t \to \infty} \tilde{\rho} (e^{-t\DG} A) = \rho_\omega (A) \, .
\end{equation}
\end{theorem}

\noindent {\bf Proof: } 1)- Assume that $\DG_\omega^\star(A)=0$. Then thanks to Proposition~\ref{mott09.prop-DGsad}, it follows that 
$0= \rho_\omega(A^\ast \DG_\omega^\star(A)) =
\sum_{x} \Gamma_{x\rightarrow \star} \rho_\omega\left(|[a_x,A]_g|^2\right) + \sum_{x} \Gamma_{\star\rightarrow x} \rho_\omega\left(|[a_x^\dag,A]_g|^2\right)$. 
Thanks to Proposition~\ref{mott09.prop-faithful}, $\rho_\omega$ is faithful, so that 
$$
[a_x,A]_g = 0 = [a_x^\dag,A]_g
\qquad
  \forall x\in\Ll(s)\,.
$$
\noindent Then the only elements of $\AG(\omega)$ with this property are the multiples of $\id$. To prove this let $A$ be a local observable first, 
namely $A\in\AG_\Lambda(\omega)$ . Then, using the decomposition into monomials $A$ can be written as $A= n_x B+(1-n_x)B' + a_x C+ a_x^\dag C'$ 
where $B,B',C,C' \in \AG_{\Lambda\setminus \{x\}}(\omega)$. Thus $[a_x,A]_g = C+a_x (B-B')=0$ implies that $C=0$ and $B=B'$. In much the same way 
$[a_x^\dag,A]_g=0$ implies $C'=0$. Thus $A=B\in\AG_{\Lambda\setminus \{x\}}(\omega)$. Using this argument
inductively on every point of $\Lambda$ shows that $A$ is a multiple of $\id$. 

\noindent If now $A\in\AG(\omega)$, 
then for all $\forall \varepsilon>0$ there is 
a finite set $\Lambda_\varepsilon$ and $A_\varepsilon\in \AG_{\Lambda_\varepsilon}(\omega)$ such that $\|A-A_\varepsilon\|<\varepsilon/3$. 
For any finite subset $\Lambda$ of $\Ll$ let $\EM_\Lambda  : \AG (\omega ) \to \AG_\Lambda (\omega )$ be the canonical projection. The commutation rule above implies that $\EM_\Lambda(A)$ also commutes with all the $a_x,a_x^\dag$ for $x\in\Lambda$, showing that there is $c(\Lambda)\in \CM$ such that $\EM_\Lambda(A)= c(\Lambda)\id$. For $\Lambda \supset \Lambda_\varepsilon$, $\|\EM_\Lambda(A-A_\varepsilon)\|= \|c(\Lambda)\id - A_\varepsilon\|<\epsilon/3$. It follows that (i) $|c(\Lambda)|\leq \|A\|$ and (ii)  $|c(\Lambda)-c(\Lambda')|<2\varepsilon/3$ for $\Lambda,\Lambda' \supset \Lambda_\varepsilon$. Thus it is a Cauchy sequence which converges to $c\in\CM$ as the volume tends to infinity. Hence $\|A-c\id\|\leq \|A-A_\varepsilon\|+ |c-c(\Lambda)| + \|A_\varepsilon- c(\Lambda)\id\|< \varepsilon$. This shows that $A=c\id$. 

\vspace{.1cm}

\noindent 2)- Assume that $\DG (A)=0$. Then 
$$
0= \rho_\omega (A^* \DG_\omega (A))= \rho_\omega (A^* \DG_\omega ^{kin} (A)) + 
\rho_\omega (A^* \DG_\omega ^\star (A))\, .
$$
By Proposition \ref{mott09.prop-DGsad} applied to $\DG_\omega^{kin}$ we obtain that
$\rho_\omega (A^* \DG_\omega ^{kin} (A)) \geq 0$. Thus 
$$
0=\rho_\omega (A^* \DG_\omega ^{kin} (A)) =
\rho_\omega (A^* \DG_\omega ^\star (A))\, .
$$
Hence by 1), $A$ is a multiple of the identity.

\vspace{.1cm}

\noindent 3)- By construction, both $\DG_\omega$ and $\DG_\omega^\star$ leave each $\AG_\Lambda(\omega)$ invariant. 
Therefore $e^{-t \DG _\omega}$ and $e^{-t\DG_\omega^\star}$ are well defined 
on each of the $\AG_\Lambda(\omega)$'s and define  Markov semi-groups (Proposition \ref{mott09.prop-CPLind}). Since the multiples 
of the identity  are the only invariant observables of the semigroups, $1$ is a simple eigenvalue of these semi-groups. 
Consequently $\lim_{t\rightarrow \infty} e^{-t\DG_\omega}(A)$ and  $\lim_{t\rightarrow \infty} e^{-t\DG_\omega^\star}(A)$ 
exist for each $A\in\AGl(\omega)$ and, (since these limits are invariant observable for the corresponding semigroups), 
these limits are multiples $m_\omega(A)$ and $m^\star_\omega (A)$ respectively of the identity. Since the semi-groups are contractions the result 
extends by continuity to all elements of $\AG(\omega)$.

\noindent Theorem~\ref{mott09.th-convAG} allows us to define $e^{-t\DG_\omega}$. 
For every $A \in \AG_\omega$, this leads to 
\begin{equation} \label{mmstar}
\lim_{t\rightarrow \infty} e^{-t\DG_\omega}(A) = m_\omega(A) \id \quad \mbox{and} \quad 
\lim_{t\rightarrow \infty} e^{-t\DG_\omega^\star }(A)= m^\star_\omega (A) \id \, .
\end{equation}

\vspace{.1cm}

\noindent 4)- By eq.~(\ref{mmstar}) it follows that $m_\omega$ and $m^\star_\omega$ are states on $\AG(\omega)$. If now $\rho$ is a 
$\DG_\omega$-invariant state and $A \in \AG_\omega$, then
$$
m_\omega(A) = \lim_{t\rightarrow \infty}
   \rho\left( e^{-t\DG_\omega}(A)\right) =
    \rho(A)\,.
$$
\noindent Similarly, if $\rho$ is a $\DG^\star_\omega$-invariant state then $m^\star_\omega = \rho$. 

\vspace{.1cm}

\noindent 5)- Thanks to Proposition~\ref{mott09.prop-DGsad}, it follows that $\rho_\omega$ is both $\DG^\star$ and $\DG$-invariant,
thus $\rho_\omega= m_\omega$ and $\rho_\omega = m^\star_\omega$.
Therefore $\rho_\omega$ is the unique invariant state for $\DG_\omega$ and $\DG_\omega^\star$.

\vspace{.1cm}

\noindent 6)- Now let $\tilde{\rho}$ be any state on $\AG_\omega$. By eq.~(\ref{mmstar}), 
$$
\lim_{t \to \infty} \tilde{\rho}\left( e^{-t\DG_\omega}(A)\right) = \tilde{\rho}\left(m_\omega (A) \id\right) = m_\omega (A)= \rho_\omega (A) 
$$
which shows the return to equilibrium $\rho_\omega$.
\hfill $\Box$

\vspace{.1cm}

\noindent As explained in Section~\ref{mott09.ssect-jumpdy} the dissipation operators, defined previously, define self-adjoint positive operators 
on the Hilbert space of the GNS representation. Some spectral properties are given by the following result

\begin{theorem}
\label{mott09.th-spectDiss}
(i) If $\ds_\omega^\star$ and $\ds_\omega$ denote the corresponding positive self-adjoint operators acting on the GNS representation, then both admit 
$0$ as a simple eigenvalue and both have a positive gap separating zero from the rest of the spectrum bounded from below by $\Gamma_\star/2$.

\vspace{.1cm}

\noindent (ii) Let $\Kk_\omega$ be the closed subspace of $\Hh_\omega$ generated by vectors in $\pi_\omega(\CG_\omega)\xi_\omega$ 
which are orthogonal to $\xi_\omega$. Then both $\ds_\omega^\star$ and $\ds_\omega$ leave $\Kk_\omega$ invariant and their restriction to $\Kk_\omega$ is bounded below by $\Gamma_\star$.

\vspace{.1cm}

\noindent (iii) The operator $\ds_\omega^\star$ has a pure point spectrum. 
\end{theorem}

\begin{rem}
\label{mott09.rem-return}
This result shows that the Markov semi-group generated by $\ds_\omega$  converges to equilibrium exponentially fast with a lifetime given
 by the inverse of $\Gamma_\star$. 
\end{rem}

\noindent {\bf Proof: } (i) In the Hilbert space $\Hh_\omega$ of the GNS representation, monomials in the annihilation-creation operators make up a total set. 
This helps creating an orthonormal basis in this Hilbert space. 
First, for $x\in\Ll$ let $b_x(\omega),b_x^\dag(\omega)$ be defined by
\begin{equation}
b_x(\omega)= e^{-\beta(\mu -\epsilon_x)^+/2}\;s_x\;a_x\,,
  \qquad
   b_x^\dag(\omega)= e^{-\beta(\epsilon_x-\mu)^+/2} \;s_x\;a_x^\dag\,.
\end{equation}
\noindent It is easy to check that both elements define unit vectors in $\Hh_\omega$, and that they are orthogonal to each other. For 
$X=(x_1,x_2\cdots, x_m)\in \Ll(\omega)^{\times m}$ let $b_X(\omega), b_X^\dag(\omega)\in\Hh_\omega$ be defined by (omitting the reference to $\omega$)
\begin{equation}
b_X= b_{x_1}b_{x_2}\cdots b_{x_m}\,,
\qquad
   b_X^\dag = b_{x_m}^\dag b_{x_{m-1}}^\dag \cdots b_{x_1}^\dag\,.
\end{equation}
\noindent If two components of $X$ are equal, then $b_X=0$. Thus the $X$'s will be restricted to the set of elements in $\Ll(\omega)^{\times n}$ made of 
distinct points. If the order in presenting the points of $X$ is changed, then $b_X$ changes sign according to the signature of the permutation. Hence, 
modulo a sign, $b_X$ depends only upon the set $\{x_1,x_2,\cdots,x_m\}\subset \Ll(\omega)$. In much the same way the vectors $\sigma_X(\omega)$
 will be defined as follows

\begin{equation}
\label{mott09.eq-sigBasis}
\sigma_x (\omega)= s_x\left(
   e^{\beta/2(\epsilon_x-\mu)}n_x - e^{-\beta/2(\epsilon_x-\mu)}(1-n_x)
                \right)\,,
\qquad
   \sigma_X = \sigma_{x_1}\sigma_{x_2}\cdots \sigma_{x_m}
\end{equation}

\noindent Here, the ordering of points is irrelevant since the $n_x$'s commute. 
Before continuing the proof we need the following lemma and corollary.

\begin{lemma}
\label{mott09.lem-GNSortho}
The vectors in $\Hh_\omega$ given by $\zeta_\omega(b_X^\dag b_Y \sigma_Z)$, 
where $X,Y,Z$ vary among the set of triplets of three disjoint finite subsets of 
$\Ll(\omega)$, including the empty set, make up an orthonormal basis of $\Hh_\omega$. 
\end{lemma}

\noindent  {\bf Proof: } First, these elements define all possible monomials in $\AG(\omega)$ up to a scalar multiplication. Consequently they generate 
a dense subspace. If it is proved that these vectors make up an orthonormal family, then they make up a Hilbert basis. Second, due to the factorization
 property of the equilibrium state $\rho_\omega$ and since $X,Y,Z$ are disjoint, it follows that 
$$\rho_\omega\left(
      (b_X^\dag b_Y \sigma_Z)^\ast\,b_{X}^\dag b_{Y} \sigma_{Z}
             \right) =
    \rho_\omega(b_Xb_X^\dag)
     \rho_\omega(b_Y^\dag b_Y)
      \rho_\omega(\sigma_Z^2)\,.
$$
\noindent Again the factorization property of $\rho_\omega$ and the commutation rules gives
$$\rho_\omega(b_Y^\dag b_Y) = 
   \prod_{i=1}^n 
    e^{-\beta|\epsilon_{y_i}-\mu|/2})\,
     \rho_\omega(n_{y_i})= 1\,.
$$
\noindent A similar result applies for $\rho_\omega(b_Xb_X^\dag)$. It also applies to $\rho_\omega(\sigma_Z^2)$ once it is remarked that 
$\sigma_x^2 = s_x\left(e^{\beta(\epsilon_x-\mu)}n_x + e^{-\beta(\epsilon_x-\mu)}(1-n_x)\right)$. Hence each of these vectors is normalized.

\vspace{.1cm}

\noindent Let now $(X,Y,Z)$ be three disjoint finite subsets of $\Ll$, and let $(X',Y',Z')$ be three other ones, such that these triples are distinct. 
Then without loss of generality, it can be assumed that $X\cup Y\cup Z \neq \emptyset$. Assume first that there is a point 
$x\in (X\cup Y\cup Z) \setminus (X'\cup Y'\cup Z')$. If $x\in X$, then the inner product
 $\rho_\omega\left((b_X^\dag b_Y \sigma_Z)^\ast\,b_{X'}^\dag b_{Y'} \sigma_{Z'}\right)$ will have a factor $\rho_\omega(b_x)=0$, if $x\in Y$, the factor 
will be $\rho_\omega(b_x^\dag)=0$, whereas if $x\in Z$ it will be $\rho_\omega(\sigma_x)=0$. Hence the two vectors are orthogonal. If 
$X\cup Y\cup Z=X'\cup Y'\cup Z'$, then either $X\neq X'$, $Y\neq Y'$ or $Z\neq Z'$. In the first case there is $x\in X\setminus X'$. Then 
$x\in Y'\cup Z'$. If $x\in Y'$, then the inner product get a factor $\rho_\omega(b_x b_x)=0$, whereas if $x\in Z'$ the factor is $\rho_\omega(b_x \sigma_x)$, 
which vanishes as well. The same argument holds in all other cases leading to the result.
\hfill $\Box$

\begin{corollary}
\label{mott09.cor-CGortho}
The family $\{\zeta_\omega(\sigma_Z)\,;\, \emptyset \neq Z\subset \Ll(\omega)\,,\, Z\;\, \mbox{\rm finite}\}$ is an orthonormal basis of $\Kk_\omega$. 
\end{corollary}

\noindent  {\bf Proof: } Since elements of $\Kk_\omega$ are orthogonal to $\xi_\omega=\zeta_\omega(\id)$ they are generated by linear combination of the
 $b_X^\dag b_Y \sigma_Z$'s with $X\cup Y \cup Z \neq \emptyset$. Since they come from the commutative sub-algebra $\CG_\omega$ they are 
generated by the operators $s_x n_x$, namely by the $\sigma_X$'s only.
\hfill $\Box$

\vspace{.2cm}

\noindent   {\bf Proof of Theorem~\ref{mott09.th-spectDiss} (continued): } 
Let $\DG_{x, \star}$ be the operator on $\AG(\omega)$ defined by
\begin{equation}
\DG_{x, \star}(A) = 
   \frac{1}{2}\big(a_x^\dag a_x A + A a_x^\dag a_x\big) - 
    a_x^\dag \sigma(A) a_x\,.
\end{equation}
\noindent In much the same way, let $\DG_{\star,x}$ be defined by
\begin{equation}
\DG_{\star,x}(A) = 
   \frac{1}{2}\big(a_x a_x^\dag A + A a_x a_x^\dag\big) - 
    a_x \sigma(A) a_x^\dag\,.
\end{equation}
\noindent It follows that if $A= A_x B$ with $A_x\in\AG_{\{x\}}$ and $B\notin \AG_{\{x\}}$ then $\DG_{x,\star}(A)=\DG_{x,\star}(A_x)B$ and 
the same holds for $\DG_{\star, x}$. Since $A_x$ can only be a linear combination of $\id\,,\ b_x\,,\,b_x^\dag\,,\, \sigma_x$, 
it is enough to consider each of these cases. 
This leads to 
$$\DG_{x,\star}(\id)=0\,, 
   \quad
    \DG_{x,\star}(b_x)=\frac{b_x}{2}\,, 
     \quad
      \DG_{x,\star}(b_x^\dag)=\frac{b_x^\dag}{2}\,, 
       \quad
        \DG_{x,\star}(\sigma_x)=2\cosh(\tfrac\beta2(\epsilon_x-\mu))n_x\,,
$$
$$\DG_{\star,x}(\id)=0,\, 
    \DG_{\star,x}(b_x)=\frac{b_x}{2},\,
      \DG_{\star,x}(b_x^\dag)=\frac{b_x^\dag}{2},\,
\DG_{\star,x}(\sigma_x)=-2\cosh(\tfrac\beta2(\epsilon_x-\mu))(\id -n_x)\,.
$$
\noindent Since $\DG^\star_\omega=\sum_{x\in\Ll(s)} \Gamma_{x\to\star} \DG_{x,\star} + \Gamma_{\star\to x} \DG_{\star,x}$  
it follows immediately  
that $b_x$ and $b_x^\dag$ are both eigenvectors of $\DG^\star_\omega$ for the common eigenvalue
\begin{equation}\label{eq-gamma_x}
 \gamma_x = \frac{\Gamma_{x\to\star}}2 + \frac{\Gamma_{\star\to x}}2 =
\frac{\Gamma_\star}{2} \left(1+e^{-\beta|\epsilon_x-\mu|} \right)\;.
\end{equation}
This means $\zeta_\omega(b_x)$ and $\zeta_\omega(b_x^\dag)$ are eigenvectors of
$\ds^\star_\omega$ with eigenvalue $\gamma_x$.
Similarly, an elementary calculation shows that
\begin{equation}
\label{mott09.eq-eigen}
\ds^\star_\omega (\zeta_\omega(b_X^\dag b_Y \sigma_Z)) =
   \gamma_{X,Y,Z} \; \zeta_\omega(b_X^\dag b_Y \sigma_Z)\,,
\end{equation}
where
\begin{equation} \label{eigenvalue}
\gamma_{X,Y,Z} =  \sum_{x\in X\cup Y} \gamma_x \,+\,
2\sum_{z\in Z} \gamma_z\;.
\end{equation}


\noindent Hence, $\ds^\star_\omega$ has pure point spectrum. In addition, $\gamma_{X,Y,Z}$ vanishes if and only if $X=Y=Z=\emptyset$. This 
shows that $\xi_\omega=\zeta_\omega(\id)$ is the only eigenvector with eigenvalues $0$. It shows also that $\Kk_\omega$ is invariant by $\ds_\omega^\star$ and, thanks to the Corollary~\ref{mott09.cor-CGortho}, the restriction of 
$\ds_\omega^\star$ on $\Kk_\omega$ is bounded below by $\Gamma_\star$ (uniformly in $\beta, \mu$ and $\omega$). 

\vspace{.1cm}

\noindent  Let
$\DG_{x,y}(A) = \frac12(L_{x\rightarrow y}^\ast L_{x\rightarrow y} A+ AL_{x\rightarrow y}^\ast L_{x\rightarrow y})- L_{x\rightarrow y}^\ast A L_{x\rightarrow y}$,
then one has
 $\DG_{x,y}(n_x) = \Gamma_{x\rightarrow y} n_x(\id -n_y),\;\DG_{x,y}(n_y)=- \Gamma_{x\rightarrow y}  n_x(1-n_y)$ and $\DG_{x,y}(n_z)=0$ for $z\not\in\{x,y\}$.
Hence, $\DG_\omega$ preserves the sub-algebra $\CG_\omega$.  
Moreover $\DG_\omega(\id)=0$, showing that $\ds_\omega$ leaves the subspace $\Kk_\omega$ invariant.  
As $\ds_\omega \geq \ds_\omega^\star$ it follows that 
$0$ is a simple eigenvalue of $\ds_\omega$ and that $\ds_\omega$ has at least the same gap as $\ds^\star_\omega$.
\hfill $\Box$

\begin{rem}
\label{mott09.rem-Gamma3}
If $\Gamma_\star$ is not constant but depends on $\epsilon_x$ and $\beta$ as specified by 
Remark~\ref{mott09.rem-Gamma2} then 
\eqref{mott09.eq-eigen} and
\eqref{eigenvalue} are still correct. 
Only the values of $\gamma_x$ would change correspondingly as one has to replace $\Gamma_\star$ by
$\Gamma_\star(\epsilon_x,\beta)$ in \eqref{eq-gamma_x}.

\noindent By the assumptions of Remark~\ref{mott09.rem-Gamma2}, the infimum of this constant for all $\epsilon_x \in \Delta$ and all $\beta \in (0, \infty )$ is positive, and therefore $\ds_\omega^\star$ and $\ds_\omega$ will still have a spectral gap (uniformly in $\beta$, $\mu$ and $\omega$) as asserted in Theorem~\ref{mott09.th-spectDiss}.
\end{rem}


\end{document}